\documentclass[12pt,aps,english]{revtex4}
\usepackage{graphicx}
\usepackage{bm}
\usepackage{dcolumn}
\usepackage[usenames, dvipsnames]{color}
\usepackage[normalem]{ulem}
\usepackage{epsfig}
\usepackage{ulem}
\usepackage{amsmath}
\usepackage{mathrsfs}

\begin{document}

\title{Translational and rotational dynamics of a self-propelled Janus probe in crowded environments}
\author{Ligesh Theeyancheri$^{1}$, Subhasish Chaki$^{1}$, Nairhita Samanta$^{1}$, Rohit Goswami$^{1}$, Raghunath Chelakkot$^{2,\dagger}$ and Rajarshi Chakrabarti$^{1, \ast}$}
\affiliation{$1.$ Department of Chemistry, Indian Institute of Technology Bombay, Mumbai, Powai 400076, E-mail: rajarshi@chem.iitb.ac.in}
\affiliation{$2.$ Department of Physics, Indian Institute of Technology Bombay, Mumbai, Powai 400076, E-mail: raghu@phy.iitb.ac.in}
\date{\today}

\maketitle

\noindent We computationally investigate the dynamics of a self-propelled Janus probe in crowded environments. The crowding is caused by the presence of viscoelastic polymers or non-viscoelastic disconnected monomers. Our simulations show that the translational, as well as rotational mean square displacements, have a distinctive three-step growth for fixed values of self-propulsion force, and steadily increase with self-propulsion, irrespective of the nature of the crowder.  On the other hand, in the absence of crowders, the rotational dynamics of the Janus probe is independent of self-propulsion force. On replacing the repulsive polymers with sticky ones, translational and rotational mean square displacements of the Janus probe show a sharp drop. Since different faces of a Janus particle interact differently with the environment, we show that the direction of self-propulsion also affects its dynamics. The ratio of long-time translational and rotational diffusivities of the self-propelled probe with a fixed self-propulsion, when plotted against the area fraction of the crowders, passes through a minima and at higher area fraction merges to its value in the absence of the crowder. This points towards the decoupling of translational and rotational dynamics of the self-propelled probe at intermediate area fraction of the crowders. However, such translational-rotational decoupling is absent for passive probes.

\section{Introduction} \label{sec:intro}
\noindent Non-equilibrium events such as force generation powered by ATP hydrolysis and directed motion at the level of individual constituents in living cells are essential to cell activities, including intracellular transport, cell motility, and cell division \cite{boal2002mechanics}. For example, oscillations of the mitotic spindle during cell division \cite{pecreaux2006spindle} or  periodic beating in flagellar motion for locomotion \cite{chelakkot2014flagellar} clearly indicates  that living (or active) matter operates far from any equilibrium state. On the other hand,  researchers have come up with innovative designs of artificial micro-swimmer \cite{patra2013intelligent,schattling2015enhanced} such as self-propelled Janus particle where self-propulsion is achieved either by coating with catalytic patches \cite{illien2017fuelled,wu2012autonomous,samin2015self} or illumination by laser light \cite{buttinoni2012active}. \\

\noindent To elucidate the role of active forces on the dynamics, in a very early experiment Wu and Libchaber have examined how a passive colloid diffuses in a bacterial bath \cite{wu2000particle}. The experimental results reveal that the imbalance between the injected energy and the dissipated heat leads to a  transient super-diffusion followed by a long-time enhanced translational diffusion. Subsequently, a series of theoretical and experimental studies have  been performed to investigate the active dynamics in thermal \cite{bechinger2016active,dhar2019run,201szamel4self,goswami2019diffusion,jee2018catalytic,chari2019scalar,lowen2019inertial,das2019active,krishnamurthy2016micrometre,das2019deviations,szamel2019stochastic,ben2011effective,redner2013structure,maggi2014generalized,ghosh2014dynamics, netz2018fluctuation,dauchot2019dynamics,chaki2018,chaki2019effects,chaki2019enhanced,samanta2016chain,eisenstecken2017internal,khalilian2016obstruction}, glassy \cite{nandi2018random,mandal2016active,berthier2019glassy,Janssen2018Mode,klongvessa2019active}  and crowded \cite{thapa2019transient,mokhtari2019dynamics,bhattacharjee2019bacterial,bhattacharjee2019confinement} environments. However, in all these studies, the persistence time of the directional motion is solely determined by thermal rotational relaxation time which is independent of activity. In contrast, a dramatic change in rotational mean square displacement (RMSD) of a self-propelled Janus has been experimentally observed in a viscoelastic environment \cite{gomez2016dynamics,lozano2018run,saad2019diffusiophoresis}. Previously analytical and computational attempts have been made to understand the dynamics of self-motile particles in a viscoelastic environment \cite{sahoo2019enhanced,vandebroek2015dynamics,du2019study,yuan2019activity,qienhanced,loi2011non}. However, understanding of the fascinating nontrivial coupling between the self-propulsion velocity and the rotational diffusivity in the presence of crowding is unresolved. A more fundamental question would be whether the results obtained from experiments \cite{gomez2016dynamics,lozano2018run,saad2019diffusiophoresis} are valid only for viscoelastic environments or it is more generic and arises from obstruction by the crowders. Such studies are extremely important as self-propelled artificial devices are used for targeted drug delivery in crowded \cite{ghosh2015non,chakrabarti2013tracer,sokolov2012models,katuri2017designing} and heterogeneous environment like biological cells \cite{barkai2012single}. \\

\noindent Motivated by recent experimental studies \cite{gomez2016dynamics,lozano2018run,saad2019diffusiophoresis,narinder2018memory,narinder2019active}, in this work we aim to investigate the dynamics of self-propelled Janus probe in a crowded medium made of polymers (Fig.~\ref{fig:scheme}(a)) or disconnected monomers, referred as colloids (Fig.~\ref{fig:scheme}(b)) and employ computer simulations to analyze the dynamics. In addition, we introduce sticky zones to each polymer or a proportional number of sticky colloids to incorporate local heterogeneity in interactions. Understanding the role of this heterogeneity has practical relevance, in the context of optimizing the search processes for targeted drug delivery \cite{katuri2017designing}. 
In a typical experimental system, the dynamics of a Janus probe is affected by various factors, including long-ranged hydrodynamic interactions with the polymer segments. However, in our model of an active system, the fluid is only providing the friction and hence there is devoid of any long-range hydrodynamic interactions. This class of active matter is termed as ``dry'' active matter \cite{marchetti2013hydrodynamics}.  This approach allows us to separately study the effect of various factors that contribute to the observed dynamical properties in experiments. A qualitative comparison with experimental observations will reveal if hydrodynamic interactions play a crucial role in determining the observed dynamical behavior of the Janus probe.
We employ computer simulations to analyze the effect of excluded volume, short-ranged sticky interactions, and fluctuations of the environment on the dynamics of the tagged Janus probe.  To highlight the relevance of the directionality of the self-propulsion in an environment with local heterogeneity, we consider two different directions of the self-propulsion, towards or away from the sticky face. Our results show that with increasing activity, there is an enhancement of rotational diffusivity irrespective of whether the medium has viscoelastic polymers or non-viscoelastic colloids. This shows that a viscoelastic medium is not necessary to show such enhanced rotation  of the tracer. We also observe that the enhancement of rotational diffusion of the self-propelled Janus probe has a non-monotonic dependence on the area fraction of the medium. Whereas the translational diffusion always decreases with the area fraction. This accounts for the decoupling of translational and rotational dynamics of the self-propelled Janus probe at intermediate crowding. Our analyses provide insights into the mechanism behind the enhancement of rotational diffusion of self-propelled probes in a complex environment in general.

\section{Simulation details} \label{sec:simulation}
\begin{figure}
\centering
\includegraphics[width=0.7\linewidth]{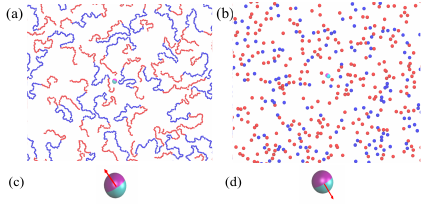} 
\caption{\small A snapshot of self-propelled Janus probe in (a) polymers with sticky zones (b) A binary mixture of colloids with sticky and non-sticky interactions (In both the cases, magenta part of the Janus probe have attractive interaction with blue colored beads). A Schematic illustration of direction of self-propulsion towards (c) sticky face (d) non-sticky face of the Janus probe.}
\label{fig:scheme}
\end{figure}
\noindent To model the viscoelastic crowders, we consider a system of polymer chains in a two-dimensional square box with periodic boundary conditions (Fig.~\ref{fig:scheme}(a)). Each polymer consists of 100 monomers, connected to the neighboring monomers by FENE potential,
 \begin{equation}
V_{\text{FENE}}\left(r \right)=\begin{cases} -\frac{k r_{\text{max}}^2}{2} \ln\left[1-\left( {\frac{r}{r_{\text{max}}}}\right) ^2 \right],\hspace{3mm} \mbox{if } r \leq r_{\text{max}}\\
\infty, \hspace{35mm} \mbox{otherwise}.
\end{cases}
\label{eq:FENE}
\end{equation}
\noindent where $r$ is the distance between two neighboring monomers in the polymer with a maximum extension of $r_{\textrm{max}}$, and $k$ is the force constant. We also consider a medium filled with disconnected monomers (Fig.~\ref{fig:scheme}(b)), to implement a non-viscoelastic crowded environment.   A pair of monomers, either free or connected, interact via repulsive  Weeks-Chandler-Andersen (WCA) potential \cite{weeks1971role}:
\begin{equation}
V_{\textrm{WCA}}(r)=\begin{cases}4\epsilon\left[\left(\frac{\sigma}{r}\right)^{12}-\left(\frac{\sigma}{r}\right)^{6}\right]+\epsilon, \mbox{if }r<2^{1/6}\sigma \\
0, \hspace{35mm} \mbox{otherwise},
\end{cases}
\label{eq:WCA}
\end{equation}
where $r$ is the separation between a pair of monomers in the medium, $\epsilon$ is the strength of the interaction, and $\sigma$ determines the effective interaction diameter. \\

 \noindent We model the Janus probe particle as a rigid body made of two spherical particles of the same size, separated by a distance $\delta$ (Fig.~\ref{fig:scheme}(c) and Fig.~\ref{fig:scheme}(d)), which is kept constant during the simulation. Each spherical particles in the Janus probe can interact differently with the monomers in the medium. If we consider a medium of non-adhesive (non-sticky) crowders, the interaction between both the particles of the Janus probe and the polymers are modeled by the WCA potential  (Eq.~\ref{eq:WCA}), where $r$ in this case is the separation between a monomer in the medium and the spherical particles of the Janus probe. We also study the effect of attractive interaction with the medium, for which we introduce a `sticky zone' in the middle of each polymer, ranging from $21^{st}$ to $70^{th}$ monomers (blue in Fig.~\ref{fig:scheme}(a)), or an equal number of attractive monomers if we consider a colloidal medium. The attractive interaction is implemented between  the `sticky' monomers and one half of the Janus probe via Lennard-Jones potential,
\begin{equation}
V_{\textrm{LJ}}(r)=\begin{cases}4\epsilon\left[\left(\frac{\sigma_{JN}}{r}\right)^{12}-\left(\frac{\sigma_{JN}}{r}\right)^{6}\right], \hspace{2mm} \mbox{if } r \leq r_{\textrm{cut}}\\
0 \hspace{31mm} , \mbox{otherwise}\\
\end{cases}
\label{eq:LJ}
\end{equation}
where $r$ is the separation between the `sticky' monomers and the attractive part of the Janus probe, $\epsilon$ is the strength of the interaction with an  interaction diameter $\sigma_{JN}$, and the Lennard-Jones cutoff length $r_{\textrm{cut}}$= $2.5$ $\sigma_{JN}$. The other (non-adhesive) half of the Janus probe, interact repulsively with all the monomers via the WCA potential (Eq.~\ref{eq:WCA}) but with  interaction diameter $\sigma_{JN}$. Also, with the non-adhesive monomers (red in Fig~\ref{fig:scheme}(a)), both the particles in the Janus probe interact via  WCA potential with the interaction diameter $\sigma_{JN}$. The energy is measured in terms of thermal energy $k_BT$, and we consider $\epsilon = 2$ according to this unit. The interaction diameter $\sigma_{JN}$ = 1.25 $\sigma$, where $\sigma$ is the unit of length. We take the Janus particle parameter $\delta = 0.6$ so that its shape anisotropy is not significant. Polymeric and colloidal systems with higher area fractions are created by introducing more polymer chains and more monomers respectively. The area fraction ($\phi$)  is defined as $\phi = \frac{N \pi \sigma^2}{4 L_x \times L_y}$ where, $N$ is the number of particles in the medium with diameter $\sigma$, and $L_x$, $L_y$ are the lengths of the simulation box. \\ 

\noindent We implement the following Langevin equation to simulate the dynamics of all the particles of our system with mass $m$ with the position $r_{i}$(t) at time t:
\begin{equation}
	m_{i}\frac{d^2 \textbf{r}_{i}(t)}{dt^2}=-\xi \frac{d \textbf{r}_{i}}{dt}- \sum_{j} \bigtriangledown V(\textbf{r}_i-\textbf{r}_j)+ {\bf f}_{i}(t)+{\bf f}^{\text{act}}
	\label{eq:langevineq}
\end{equation} 
	Here $r_j$ represents the position of all the particles except the $i^{th}$ particle, $V(r) = V_{\text{LJ}}+V_{\text{WCA}}+V_{\text{FENE}}$, is the resutent pair potential between the $i^{th}$ and $j^{th}$ particle. Note that  $V_{\text{LJ}}=0$ for purely repulsive interactions, $V_{\text{WCA}}=0$ for attractive interactions, and $V_{\text{FENE}}=0$ for free colloids in the environment. $\xi $ is the friction coefficient which is related to the Gaussian noise through
\begin{equation}
	\left<f_\alpha(t)\right>=0, \hspace{5mm}
	\left<f_{\alpha}(t^{\prime})f_{\beta}(t^{\prime\prime})\right> =4 \xi k_B T \delta_{\alpha\beta}\delta(t^{\prime}-t^{\prime\prime})
	\label{eq:random-forcerouse}
\end{equation}
 \noindent The term ${\bf f}^{\text{act}}$ denotes the self-propulsion term of active particles. For Janus particles,  ${\bf f}^{\text{act}}_i ={F}\hat{n}$ in Eq.~\ref{eq:langevineq}, where, $F$ is the strength of the propulsion force and $\hat{n}$ is the unit connecting vector between the two centers of the Janus probe. For passive monomers in the crowded environment, we set $F=0$.  Introduction of self-propulsion force term ensures the violation of the fluctuation-dissipation theorem (FDT)\cite{bechinger2016active, chaki2019effects, chaki2018} shown in Eq.~\ref{eq:random-forcerouse}. All the simulations are performed using Langevin thermostat and equation of motion integrated using the velocity Verlet algorithm in each time step. All the production simulations are carried out for $10^{7}$ steps where the integration time step is considered to be $10^{-5}$. The simulations are carried out using LAMMPS \cite{plimpton1995fast}, a freely available open-source molecular dynamics package.

\section{Results and discussion}

\begin{figure}
	\centering
	\begin{tabular}{cc}
	        \includegraphics[width=0.5\textwidth]{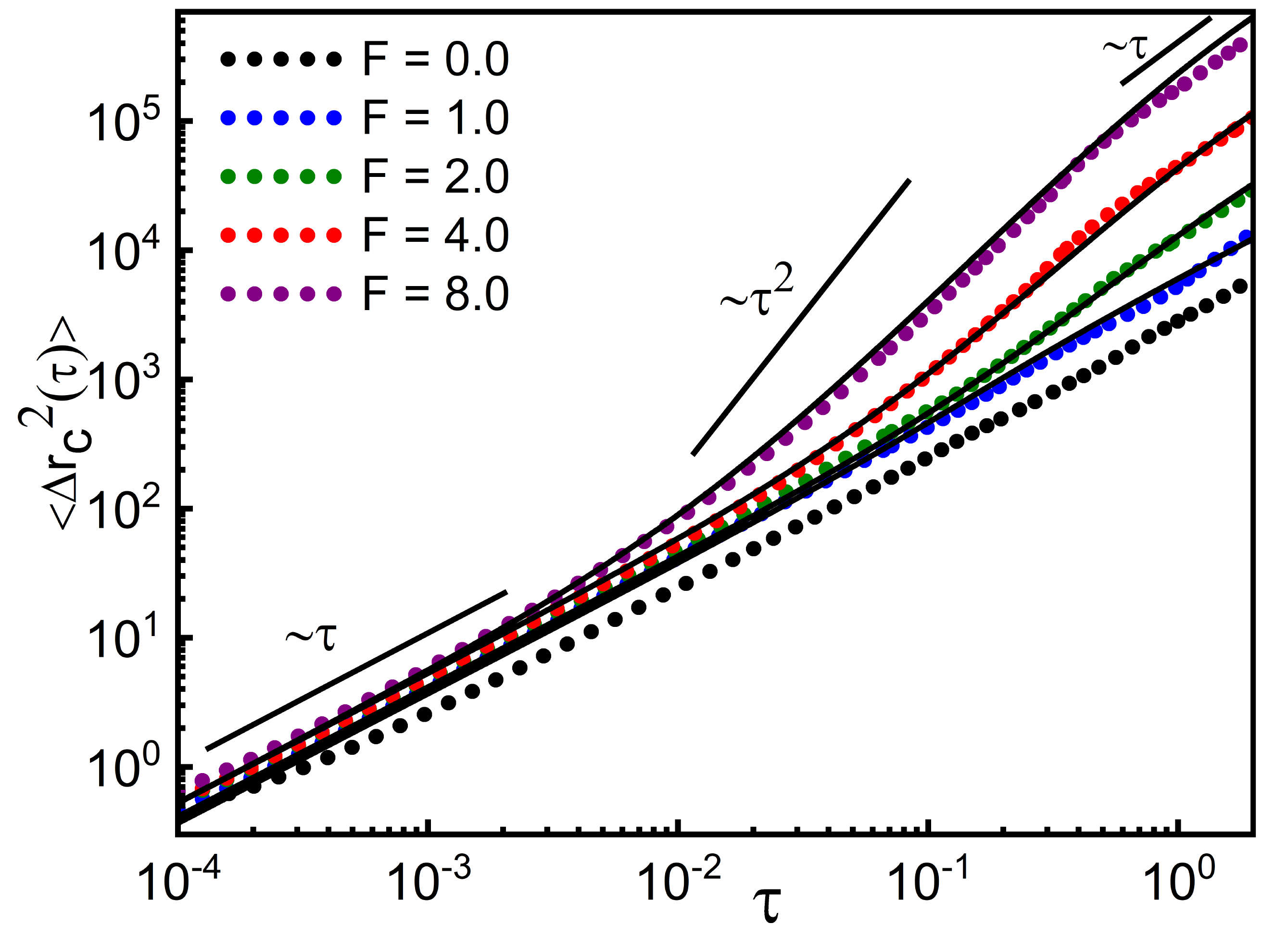}&
		\includegraphics[width=0.5\textwidth]{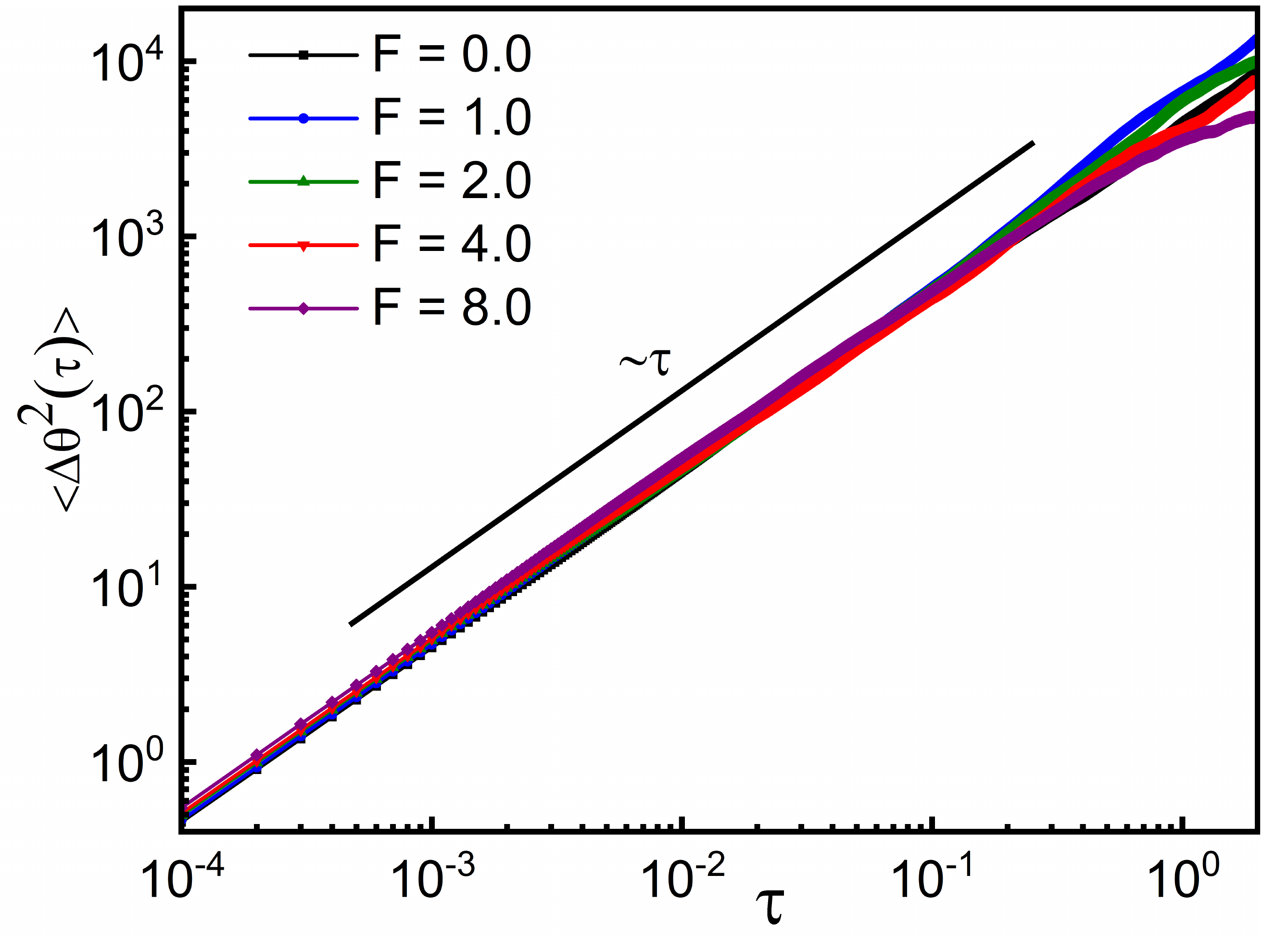} \\		
		(a) & (b)  \\
		
	\end{tabular}
	\caption{\small Log-Log plot of (a) $\left<\Delta r_c^2(\tau)\right>$ $vs$ $\tau$ for the passive (black dotted line) and self-propelled free Janus probe for different $F$ fitted with analytical expression (Eq.~\ref{eq:MSD_Analytical}) (black solid lines) (b) $\langle \Delta \theta^{2} (\tau) \rangle$ vs $\tau$ for the passive (black solid line) and self-propelled free Janus probe for different $F$.}
\label{fig:MSD_fj}
\end{figure}

\noindent  As a first step, we perform the simulation and study the dynamics of  a free Janus particle in the absence of crowders, for future comparison with its dynamics in the crowded medium. We calculate  the time-and-ensemble averaged translational mean square displacement (MSD) $\left(\left<\Delta r_c^2(\tau)\right>\right)$, and the rotational mean square displacement (RMSD) $\left(\left<\Delta \theta^2(\tau)\right>\right)$ of the Janus probe for different values of $F$ (Fig.~\ref{fig:MSD_fj}). Here $r_c$ is the centre-of-mass position of the Janus particle, and  $\theta$ is the angle between the orientation vector $\hat{n}$ and the positive $x$-axis. \\

\noindent We compare this numerically calculated MSD to the $\left<\Delta r_c^2(\tau)\right>$ curves with the analytical expression for the active Brownian particle \cite{bechinger2016active}: 
\begin{equation}
\left<\Delta r_c^2(\tau)\right> = \left[4D_{T} + 2 v^{2} \tau_{R} \right] \tau + 2 v^{2} \tau_{R}^{2} \left[ e^{-\frac{\tau}{\tau_{R}}} - 1 \right],
	\label{eq:MSD_Analytical}
\end{equation} 

\noindent where $D_T$ is the thermal translational diffusion coefficient, $v$ is the self-propelled velocity and $\tau_R $ is the persistence time of the active particle which is the inverse of the rotational diffusion coefficient, $D_R$. From Fig.~\ref{fig:MSD_fj}(a), it is evident that the translational MSD has three distinct regions: $\left<\Delta r_c^2(\tau)\right>=4D_T \tau$, when $\tau<\tau_R$; $\left<\Delta r_c^2(\tau)\right>\simeq4D_T\tau+v^2 \tau^2$ when $\tau\simeq\tau_R$; and $\left<\Delta r_c^2(\tau)\right>\simeq(4D_T+2v^2\tau_R) \tau$, when $\tau>\tau_R$ and the fitting is done with substituting $D_T$, $D_R$ obtained from the plot (Fig.~\ref{fig:MSD_fj}) in Eq.~\ref{eq:MSD_Analytical} (see ESI for more details). {We note that Eq.~\ref{eq:MSD_Analytical} holds for a two dimensional free self-propelled spherical probe (Fig.~\ref{fig:MSD_fj}(a))}. From Fig.~\ref{fig:MSD_fj}(a), one can see that in the absence of any crowders in the medium, the $\left<\Delta r_c^2(\tau)\right>$  of a self-propelled free Janus grows faster compared to the passive case. {On increasing $F$, the growth of $\left<\Delta r_c^2(\tau)\right>$ becomes increasingly faster}, while the $\left<\Delta \theta^2(\tau)\right>$ remains unaffected as the rotational motion is solely governed by thermal fluctuations. Hence the persistence time, $\tau_R$ is the same for all the curves of different $F$, when crowders are not present in the medium.

\subsection{Dynamics of self-propelled Janus probe in the presence of viscoelastic crowders (polymers)}
\begin{figure}
	\centering
	\begin{tabular}{cc}
		\includegraphics[width=0.5\textwidth]{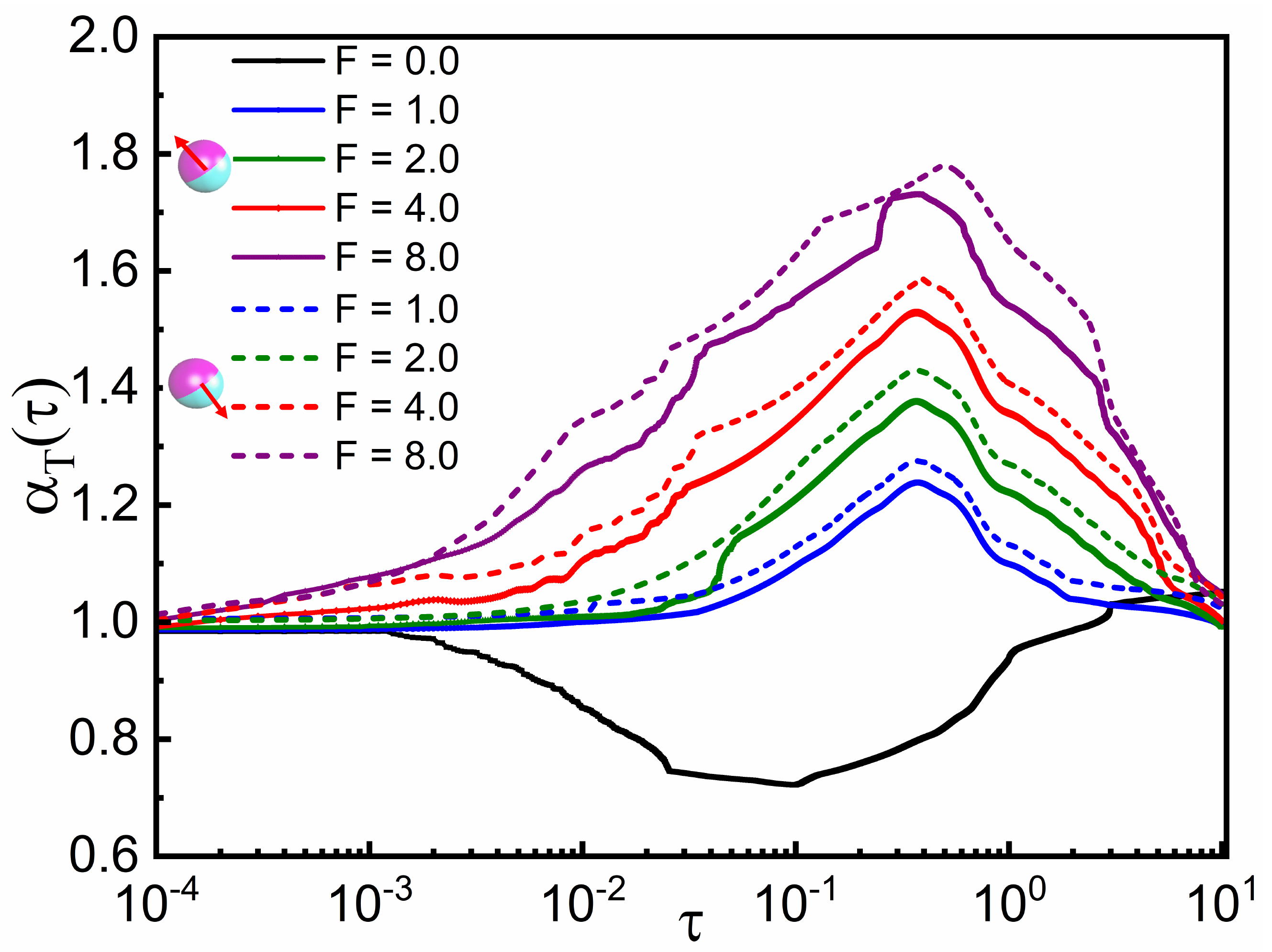} &
		\includegraphics[width=0.5\textwidth]{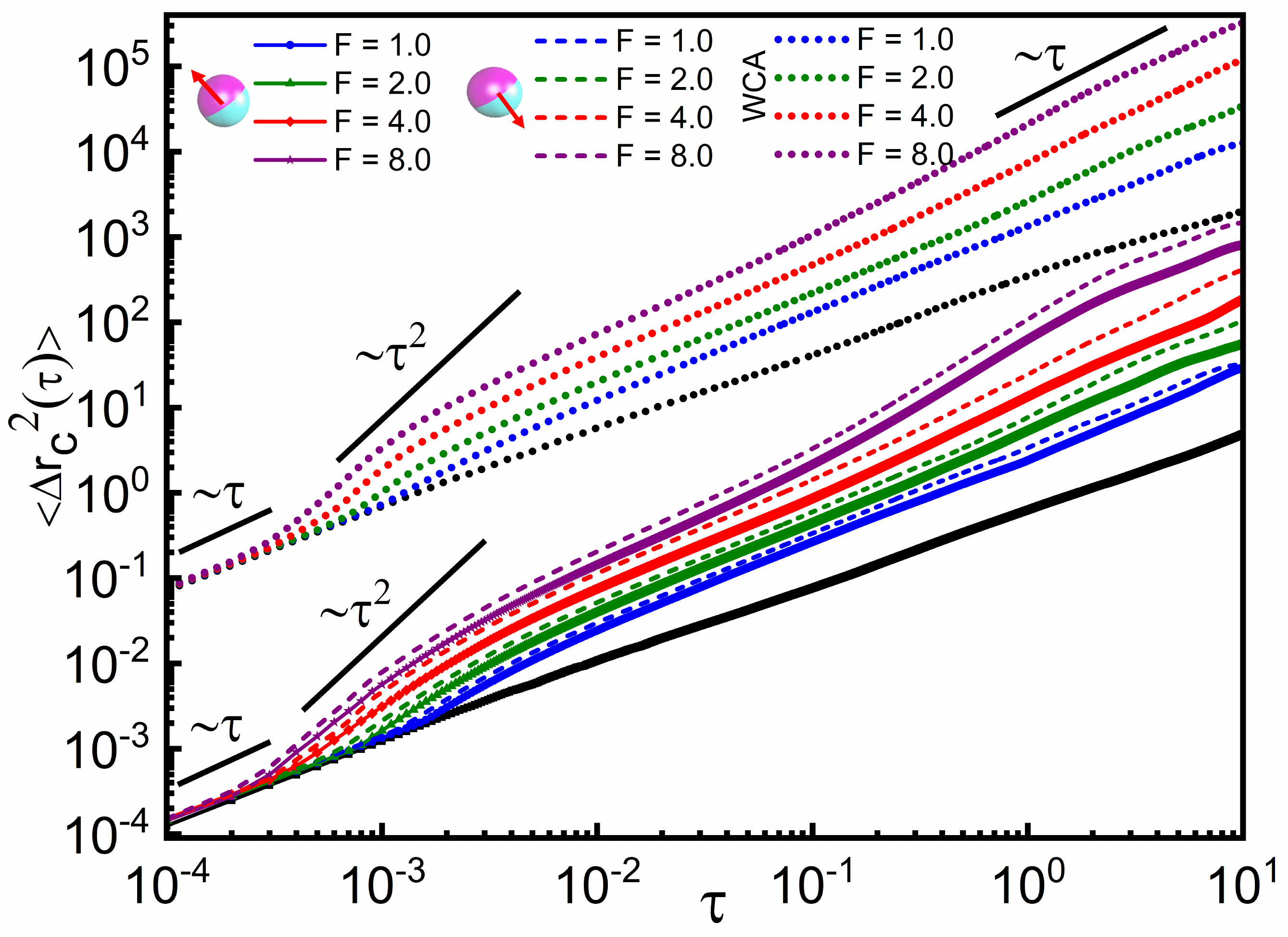} \\
		
		(a) & (b)  \\
		
	\end{tabular}
	\caption{\small (a) The time exponent $\alpha_{T} (\tau)$ for the Janus probe subjected to different self-propulsion $F$ in polymers having sticky zones ($\epsilon = 2.0$) for $\phi = 0.165$. (b) Log-Log plot of $\left<\Delta r_c^2(\tau)\right>$ $vs$ $\tau$  for the self-propelled Janus probe subjected to different $F$ in polymers having sticky zones ($\epsilon = 2.0$) for $\phi = 0.165$. The black lines (solid and dotted) represent the case for the passive Janus probe. The dotted lines represent the case of Janus probe in purely repulsive polymers. In the plots, the solid and dashed lines represent the case with self-propulsion towards the sticky face and non-sticky face respectively.}
\label{fig:TMSD_exp}
\end{figure}
\begin{figure}
	\centering
	\begin{tabular}{cc}
		\includegraphics[width=0.5\textwidth]{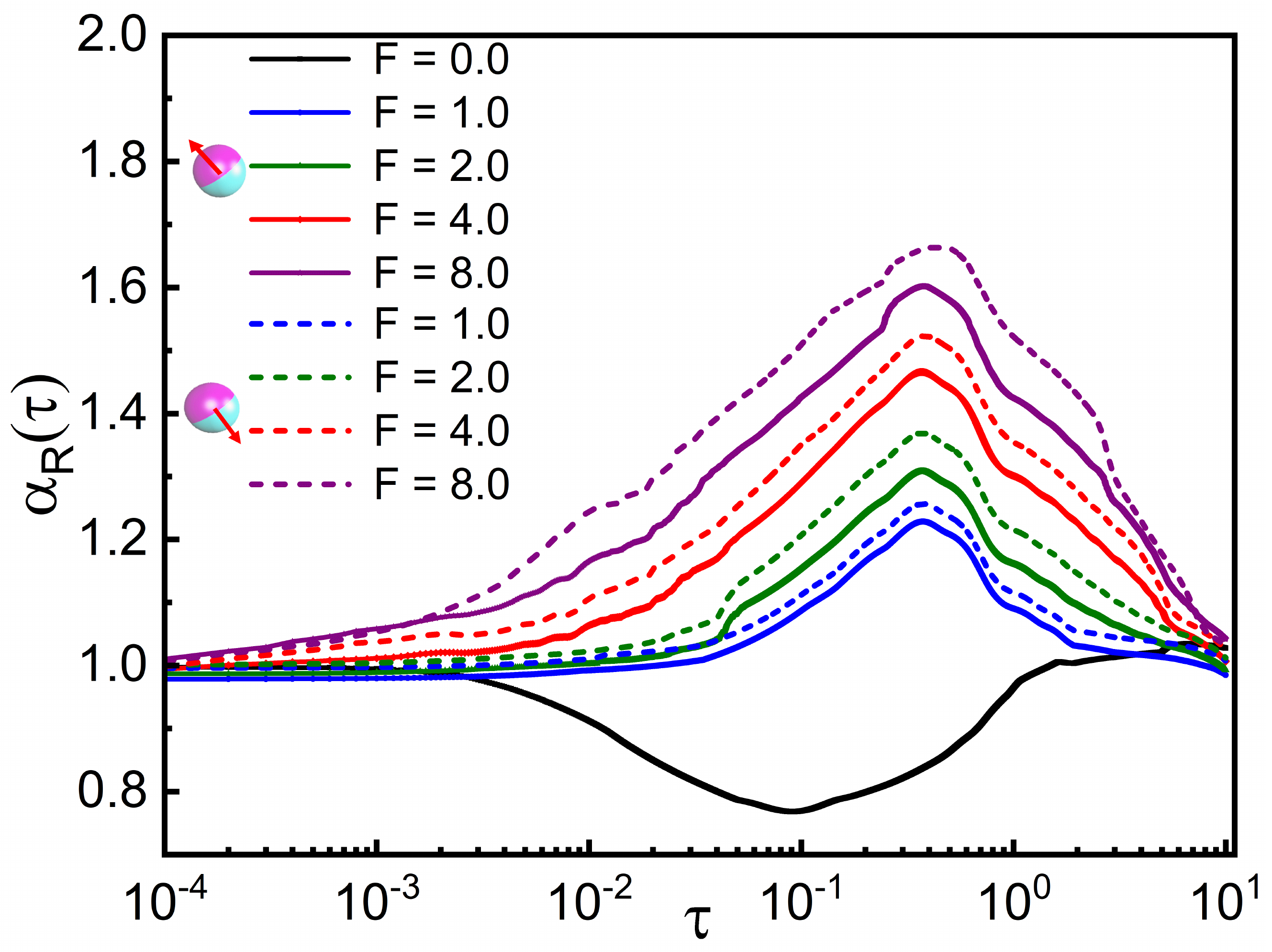} & 
		\includegraphics[width=0.5\textwidth]{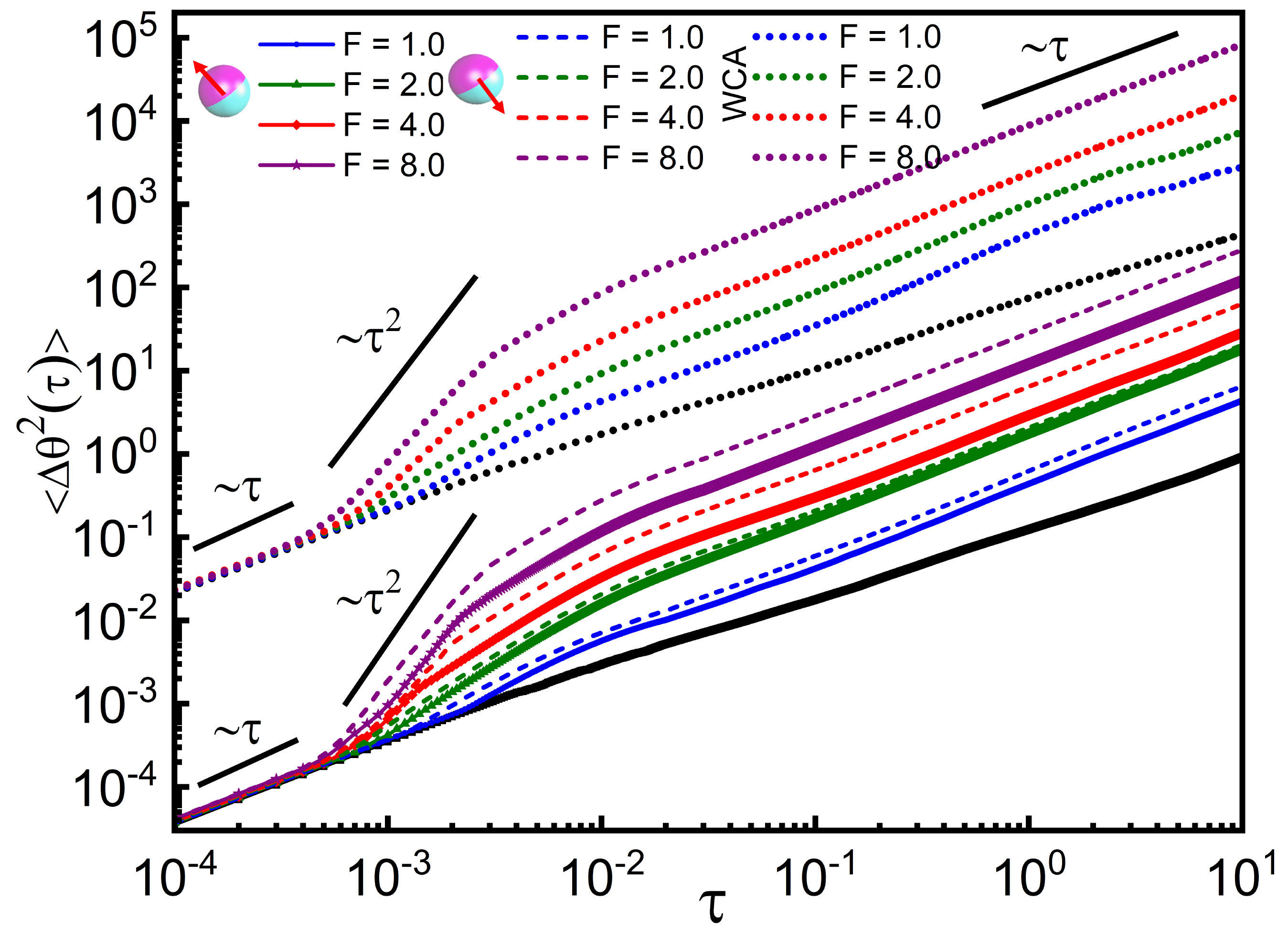} \\ 
		
		(a) & (b)  \\
		
	\end{tabular}
	\caption{\small (a) The time exponent $\alpha_{R} (\tau)$ for the Janus probe subjected to different self-propulsion $F$ in polymers having sticky zones ($\epsilon = 2.0$) for $\phi = 0.165$. (b) Log-Log plot of $\langle \Delta \theta^{2} (\tau) \rangle$ vs $\tau$ for the self-propelled Janus probe subjected to different $F$ in polymers having sticky zones ($\epsilon = 2.0$) for $\phi = 0.165$. The black lines (solid and dotted) represent the case for the passive Janus probe.The dotted lines represent the case of Janus probe in purely repulsive polymers. In the plots, the solid and dashed lines represent the case with self-propulsion towards the sticky face and non-sticky face respectively.}
\label{fig:MSAD_exp}
\end{figure}

\noindent Next, we introduce viscoelastic crowders (polymers), both adhesive and non-adhesive, in the environment and study their influence on the dynamics of the Janus probe. We first focus on the effect of interaction between the probe and the crowders, thus we keep the area fraction of the polymers to $\phi = 0.165$. When the polymers are purely repulsive they transiently surround the Janus probe, causing a sub-diffusive behavior ($(\alpha_T(\tau)<1)$ ) at intermediate time and a free diffusion $(\alpha_T(\tau)=1)$ at large times (ESI Fig.~1(a)) \cite{samanta2016tracer}, where the exponent $\alpha_T (\tau)=\frac{d \log \left<{\Delta r_c^{2}(\tau)}\right>}{d \log \tau}$. However, for an active Janus particle, the self-propulsion helps the Janus probe to escape the steric {barrier} created by these polymers, and $\left<\Delta r_c^2(\tau)\right>$ shows a three-step growth: short time thermal diffusion $(\alpha_T(\tau)=1)$, intermediate superdiffusion $(\alpha_T(\tau)>1)$ and enhanced translational diffusion at large times $(\alpha_T(\tau)=1)$ (ESI Fig.~1(a)) in comparison to the passive Janus probe in polymers (Fig.~\ref{fig:TMSD_exp}(b)). Thus, self-propulsion turns intermediate-time subdiffusion to superdiffusion. Interestingly, in the presence of polymers,  $\left<\Delta \theta^2(\tau)\right>$ of the self-propelled Janus probe  also exhibits three-step growth unlike a free self-propelled Janus probe (Fig.~\ref{fig:MSAD_exp}(b)). Similar to $\left<\Delta r_c^2(\tau)\right>$, we again observe short time thermal diffusion $(\alpha_R(\tau)=1)$, intermediate superdiffusion $(\alpha_R(\tau)>1)$ and long time enhanced free diffusion $(\alpha_R(\tau)=1)$ for $\left<\Delta \theta^2(\tau)\right>$ (Fig.~1(b)), where, $\alpha_{R} (\tau)=\frac{d \log \left<{\Delta \theta^{2}(\tau)}\right>}{d \log \tau}$. On increasing the propulsion force $F$, long-time value of $\left<\Delta \theta^2(\tau)\right>$ shows steady increase (Fig.~\ref{fig:MSAD_exp}(b)), which indicates that the Janus probe with higher self-propulsion has faster rotational dynamics even for a smaller probe size in comparison to the average size of the chains. The intermediate superdiffusive rotational dynamics of the Janus probe in the presence of crowders goes hand in hand with its translational dynamics, where the superdiffusion sets in at earlier time compared to the case with no crowders (Fig.~\ref{fig:MSD_fj}(a), ~\ref{fig:TMSD_exp}(b)). A similar trend in rotational dynamics has been observed by Gomez-Solano \textit{et. al} in an experiment, where they notice an enhanced diffusion in $\langle \Delta \theta^{2} (\tau) \rangle$ for a self-propelled Janus probe in a viscoelastic environment \cite{gomez2016dynamics}, when compared with a passive Janus in the same viscoelastic medium. However, this study do not report the three distinct dynamical regimes that we observe in the simulations. In the experiment, the Janus probe was orders of magnitude bigger than the average size of the polymeric chains. We see qualitatively similar behavior even for a smaller probe size compared to the average size of the chain. \\

\begin{figure}
	\centering
	\begin{tabular}{cc}
		\includegraphics[width=0.5\textwidth]{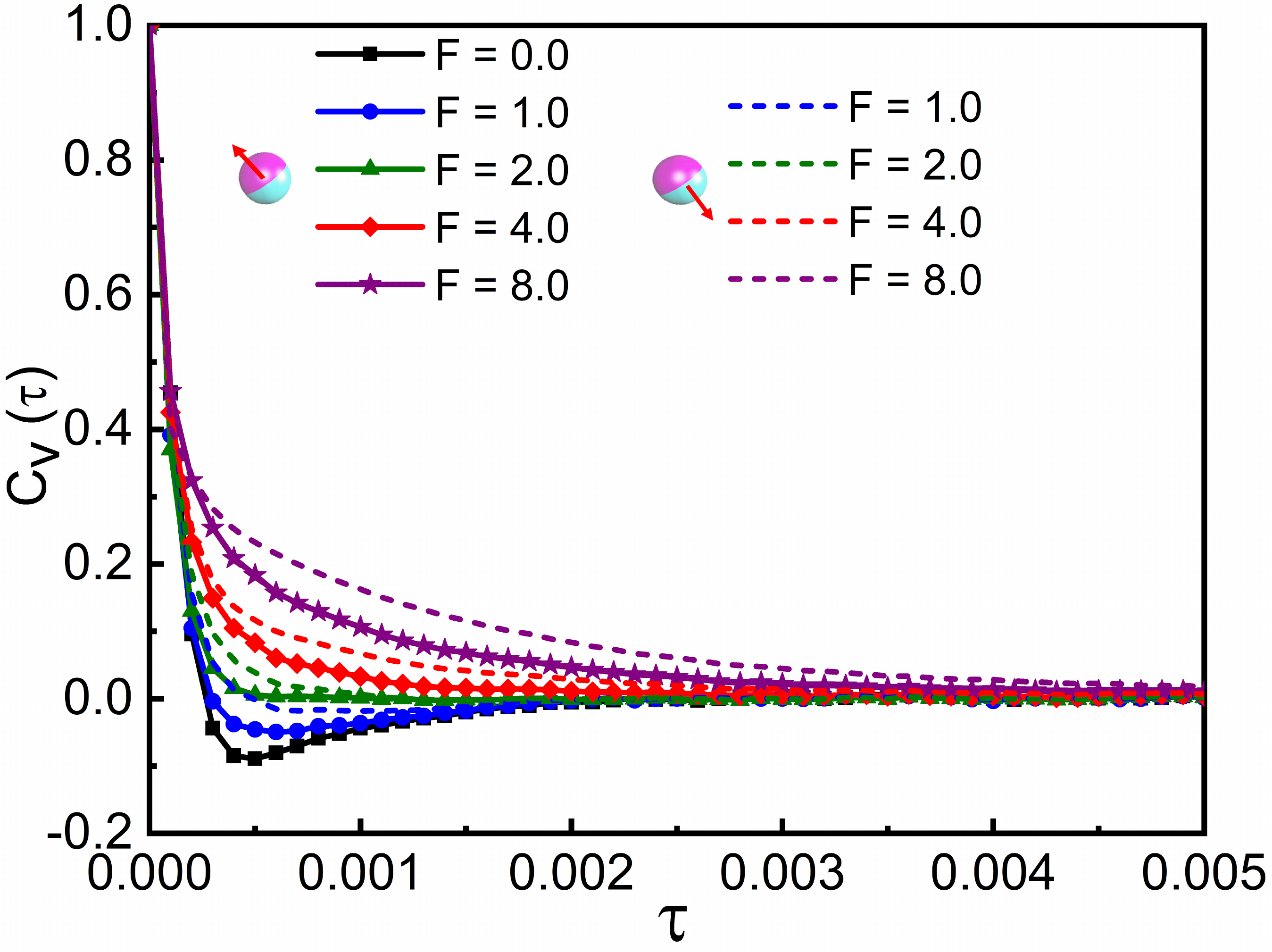} &
		\includegraphics[width=0.5\textwidth]{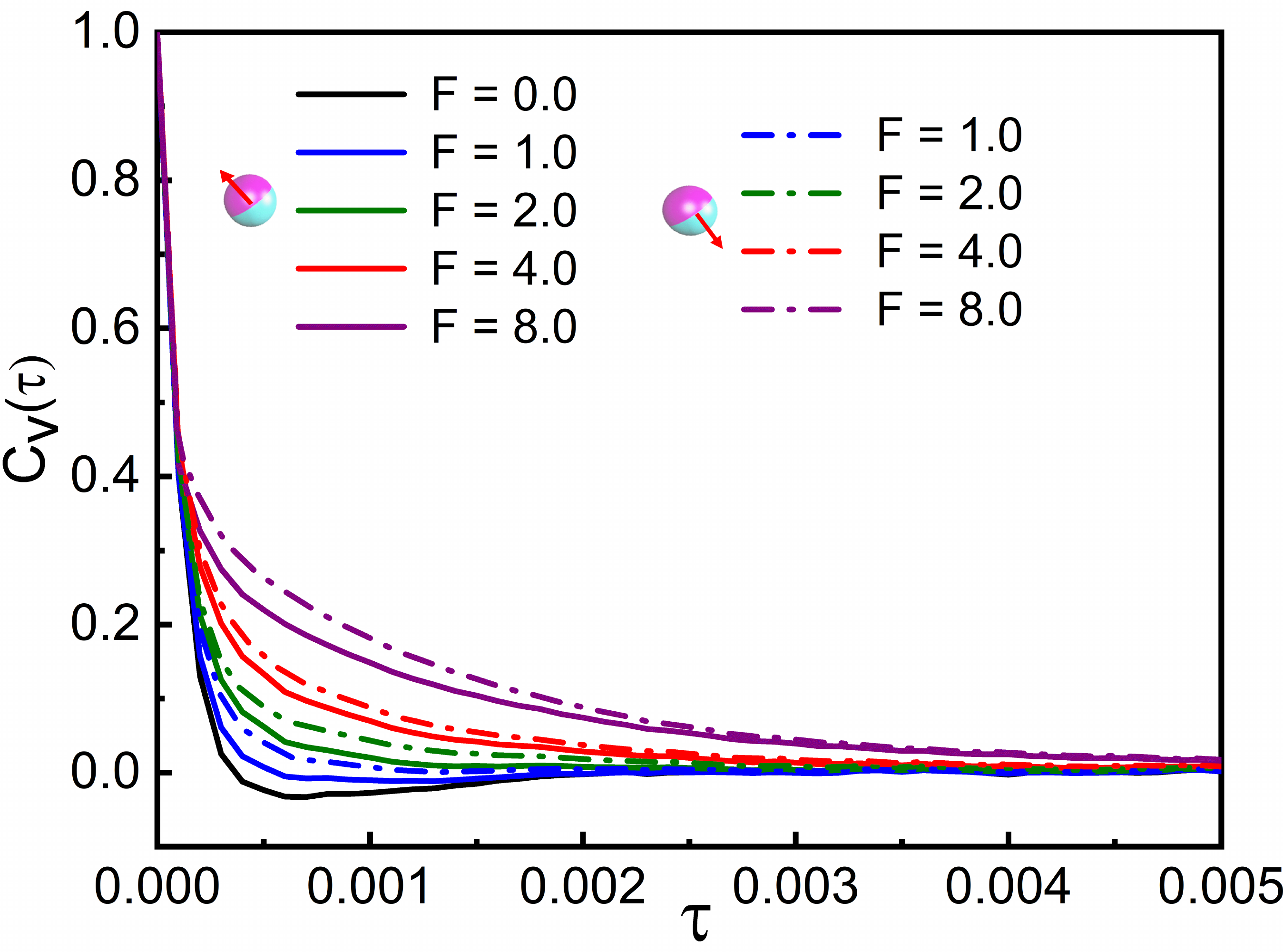} \\
		
		(a) & (b)  \\
		
	\end{tabular}
        \caption{\small The velocity autocorrelation function for the Janus probe subjected to different self-propulsion $F$ for $\phi = 0.165$ towards two different directions in (a) polymers having sticky zones ($\epsilon = 2.0$) and (b) binary mixture of colloids with sticky (LJ) ($\epsilon = 2.0$) and repulsive (WCA) interactions. The black line represents the same for the passive Janus probe.}
\label{fig:VACF_1}
\end{figure}

\noindent  Next, we incorporate local heterogeneity by replacing the repulsive polymers with the polymers having sticky zones. Thus, the probe when in proximity to a polymer chain can have sticky, as well as non-sticky partners at the same time. In the presence of sticky zones, $\left<\Delta r_c^2(\tau)\right>$ and $\left<\Delta \theta^2(\tau)\right>$ show qualitative trends similar to the case of repulsive polymers (Fig.~\ref{fig:TMSD_exp}(b), \ref{fig:MSAD_exp}(b)). However, the Janus probe is now pulled in by the sticky zones of the polymers which leads to an overall decrease in $\left<\Delta r_c^2(\tau)\right>$ and $\left<\Delta \theta^2(\tau)\right>$ in comparison to the case with repulsive polymers. Since the local heterogeneity persists {within a length-scale comparable to the probe-size}, the direction of propulsion of the Janus prob relative to its sticky face plays a crucial role in controlling the dynamics. In order to study this, we choose two different directions of the self-propulsion: a) one towards the sticky face (Fig.~\ref{fig:scheme}(c))  and  b) another towards the non-sticky face (Fig.~\ref{fig:scheme}(d)) of the Janus probe \cite{buttinoni2012active}. The direction of self-propulsion towards the non-sticky face facilitates faster translation and rotation compared to the case where the  direction of self-propulsion points towards the sticky face,  as evident from Fig.~\ref{fig:TMSD_exp} and Fig.~\ref{fig:MSAD_exp}. In other words, case a) adds to the stickiness, and case b) helps in escaping from the sticky traps \cite{wexler2020dynamics,caprini2019active}. To elucidate the probe dynamics further, we compute the velocity autocorrelation function ($C_\textrm{v}(\tau)$) for sticky, as well as non-sticky crowders. Also, for the sticky crowders, we compare  $C_\textrm{v}(\tau)$ for probes with both types of leading faces (Fig.~\ref{fig:VACF_1}). At early times $(\tau<\tau_R)$, all the $C_\textrm{v}(\tau)$s show sharp fall owing to the overdamped nature of the probe dynamics, be it passive or active.  Also, on increasing the activity, $C_\textrm{v}(\tau)$s show a stronger correlation with a higher decay time, which is more clearly visible for the probe in repulsive crowders (see ESI, Fig.~2 and Fig.~3). However, in the case of sticky crowders,  the $C_\textrm{v}(\tau)$s show negative dips for passive and smaller activities (self-propulsion forces) due to  effective trappings [see ESI, Movie-1] of the Janus probe by the polymers \cite{kumar2019transport,samanta2016tracer}. While increasing the activities, the $C_\textrm{v}(\tau)$ remain throughout positive reflecting the escaping events of the Janus probe from the traps formed by the polymers (Fig.~\ref{fig:VACF_1}) [see ESI, Movie-2]. Interestingly, for the same $F$, probes with a non-sticky leading face shows a less negative value compared to the probes with sticky leading face, as the direction of self-propulsion affects the escaping from the sticky zones [see ESI, Movie-3] as shown in Fig.~\ref{fig:VACF_1}. This behavior is consistent with the observed translational dynamics manifested in the MSD behavior in Fig.~\ref{fig:TMSD_exp}. \\
\begin{figure}
	\centering
	\begin{tabular}{cc}
		\includegraphics[width=0.5\textwidth]{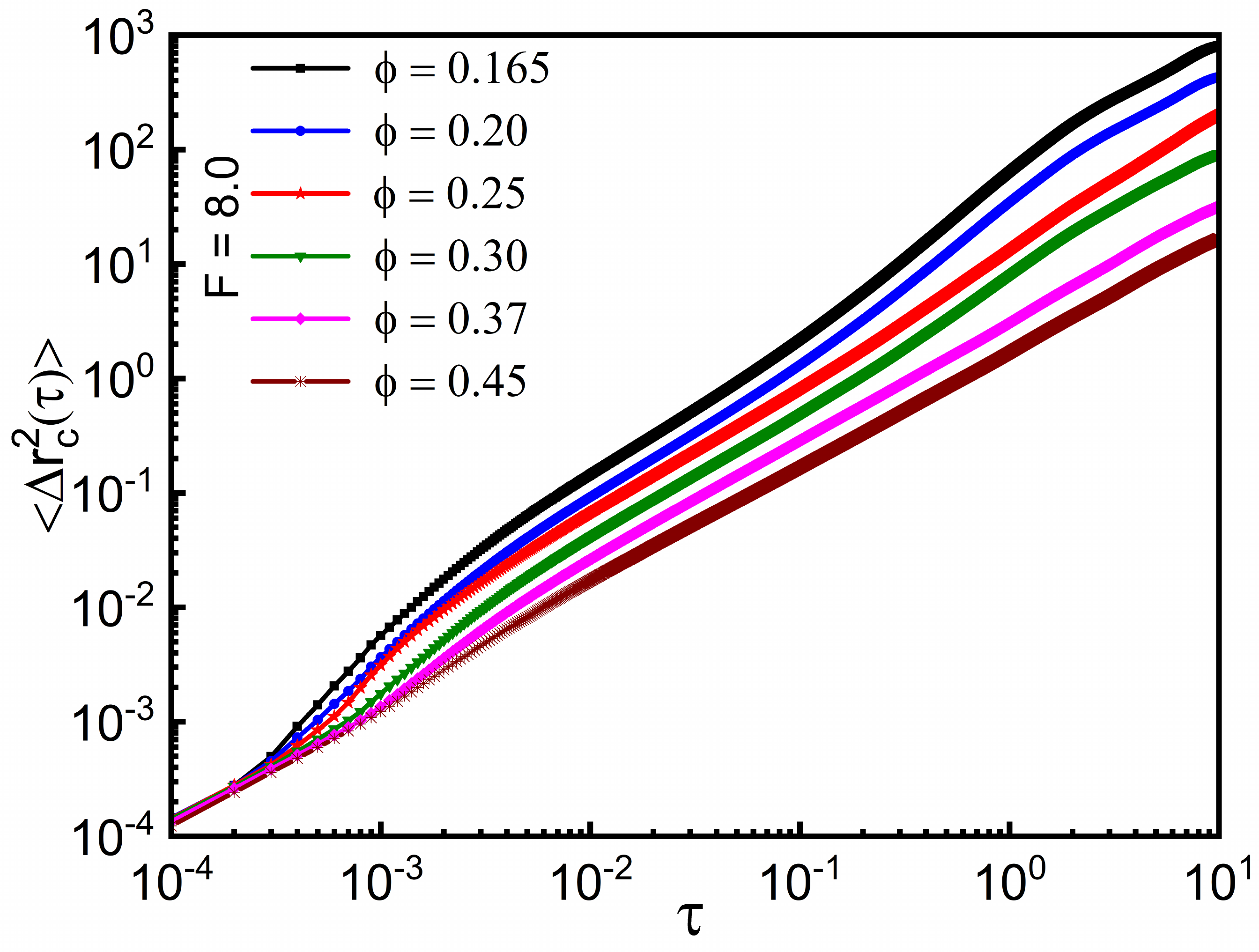} &
		\includegraphics[width=0.5\textwidth]{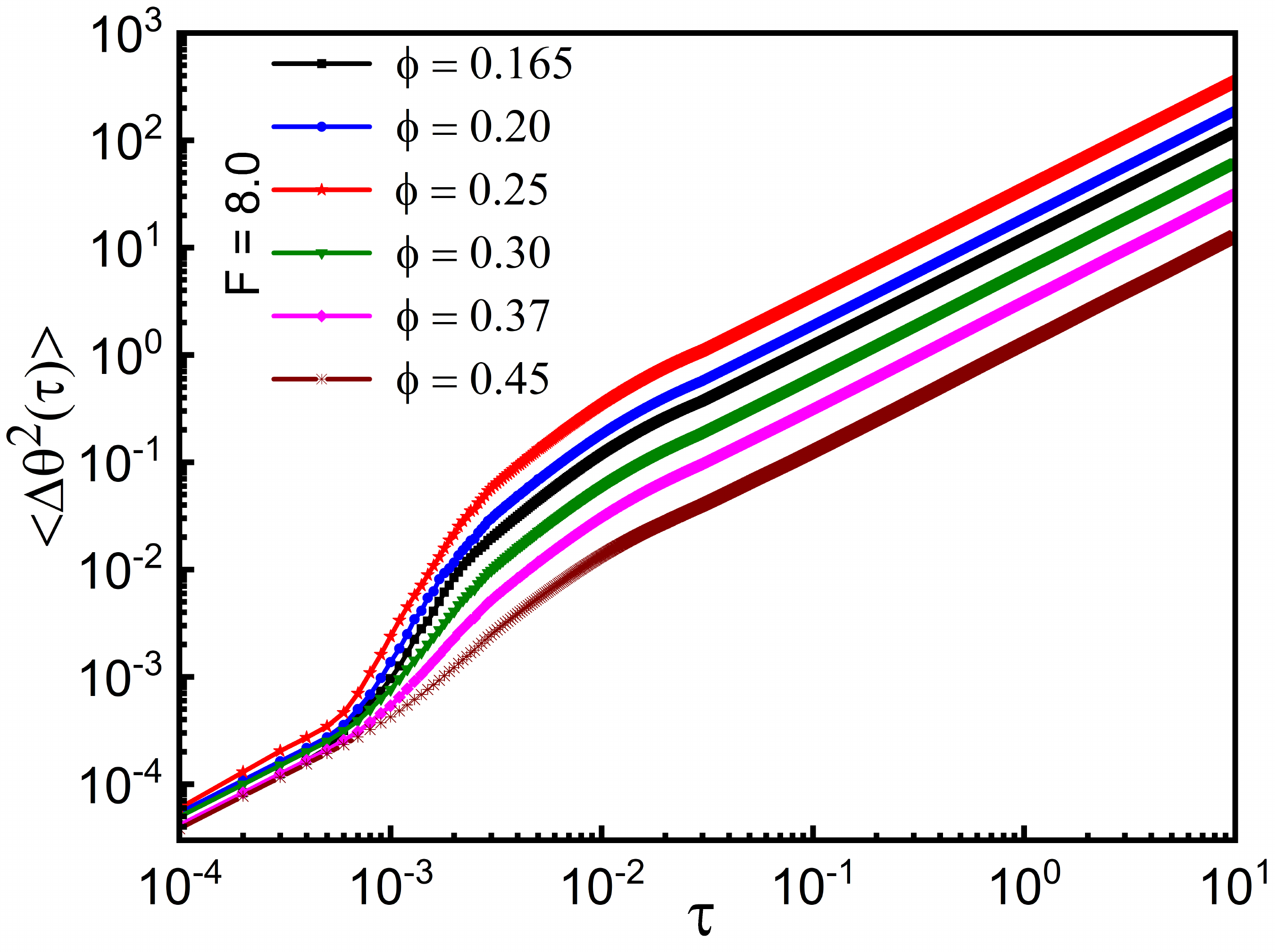} \\

                 (a) & (b)  \\
		
	\end{tabular}
              \caption{\small Log-Log plot of (a) $\left<\Delta r_c^2(\tau)\right>$ $vs$ $\tau$ (b) $\left<\Delta \theta^{2} (\tau) \right>$ $vs$ $\tau$ for the Janus probe in polymers having sticky zones ($\epsilon = 2.0$) with self-propulsion towards the sticky face in different area fraction of polymers $\phi$ for a fixed self-propulsion force $F=8.0$.}
\label{fig:MSD_phi}
\end{figure}

\noindent Subsequently, we investigate the effect of density of the medium by varying $\phi$ from $0.165$ to $0.45$, and keeping $F$ a constant. We calculate  $\left<\Delta r_c^2(\tau)\right>$ (Fig.~\ref{fig:MSD_phi}(a)) and $\left<\Delta \theta^2(\tau)\right>$ (Fig.~\ref{fig:MSD_phi}(b)) as a function of  time under a constant activity $F=8.0$. The $\left<\Delta r_c^2(\tau)\right>$ for the Janus exhibits a three-step growth with $\tau$ for all values of $\phi$. However the long-time diffusive behavior slows down with increasing $\phi$ (Fig.~\ref{fig:MSD_phi}(a)) due to an increase in crowding in the medium. Like $\left<\Delta r_c^2(\tau)\right>$, $\left<\Delta \theta^2(\tau)\right>$ also exhibits a three-step growth for $\phi \geq 0$.
However, the long-time diffusive behavior of the angular part, $\left<\Delta \theta^2(\tau)\right>$ shows a non-monotonic behavior with $\phi$ (Fig.~\ref{fig:MSD_phi}(b)). We observe an increase in $\left<\Delta \theta^2(\tau)\right>$ at $\tau \rightarrow \infty$ with $\phi$, up to $\phi = 0.25$. Further increase in $\phi$ shows a decrease in this value. This intriguing behavior is discussed in detail in section~\ref{non_monotonous}.\\

\noindent We further analyze the effect of probe-size in its dynamics by computing its  MSD varying its interaction radius $\sigma_{JN}$ for $F = 8.0$ with the crowders ($\phi = 0.165$) by keeping $\delta$ constant. An increase in $\sigma_{JN}$ changes only the effective interaction radius between the probe and the crowders while keeping effective viscous dissipation invariant. This allows us to separately study how the effect of crowding changes with the interaction radius of the Janus probe. In ESI, Fig.~4, we plot the $\left<\Delta r_c^2(\tau)\right>$ and $\langle \Delta \theta^{2} (\tau) \rangle$  for $\sigma_{JN} = 1.25, 2, 3$, which clearly indicates the slowing down of both translational and rotational dynamics of the probe with its size.

\subsection{Comparative dynamics of self-propelled Janus probe in polymers and colloids}
\begin{figure}
\centering
\includegraphics[width=0.6\linewidth]{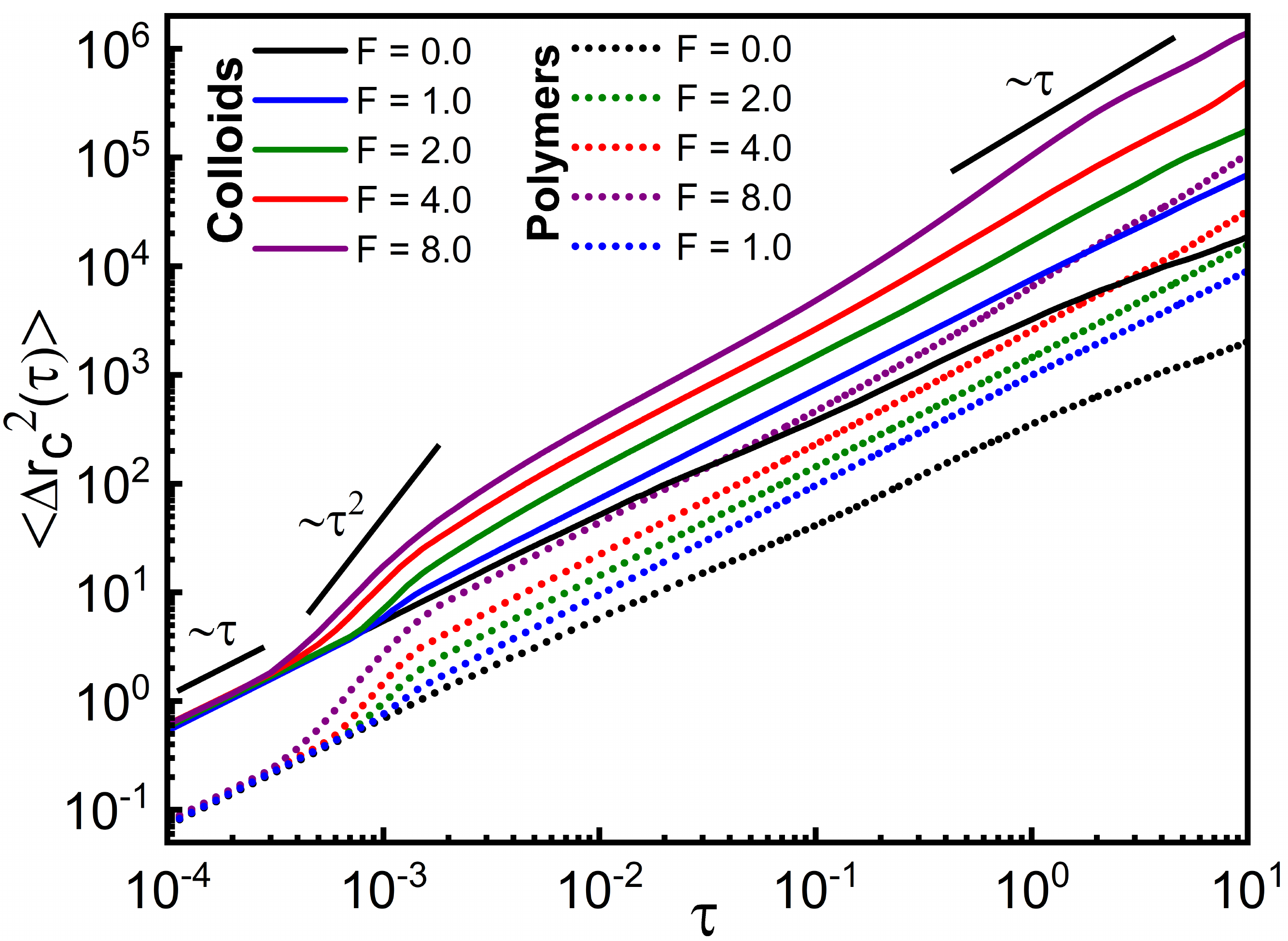} 
\caption{\small Log-Log plot of $\left<\Delta r_c^2(\tau)\right>$ $vs$ $\tau$ for the self-propelled Janus probe in the purely repulsive polymeric environment (dotted lines) and in purely repulsive colloids (solid lines) for $\phi = 0.165$. The black solid and dotted lines represent the case for the passive Janus probe.}
\label{fig:TMSD_WCA}
\end{figure}

\begin{figure}
	\centering
	\begin{tabular}{cc}
		\includegraphics[width=0.5\textwidth]{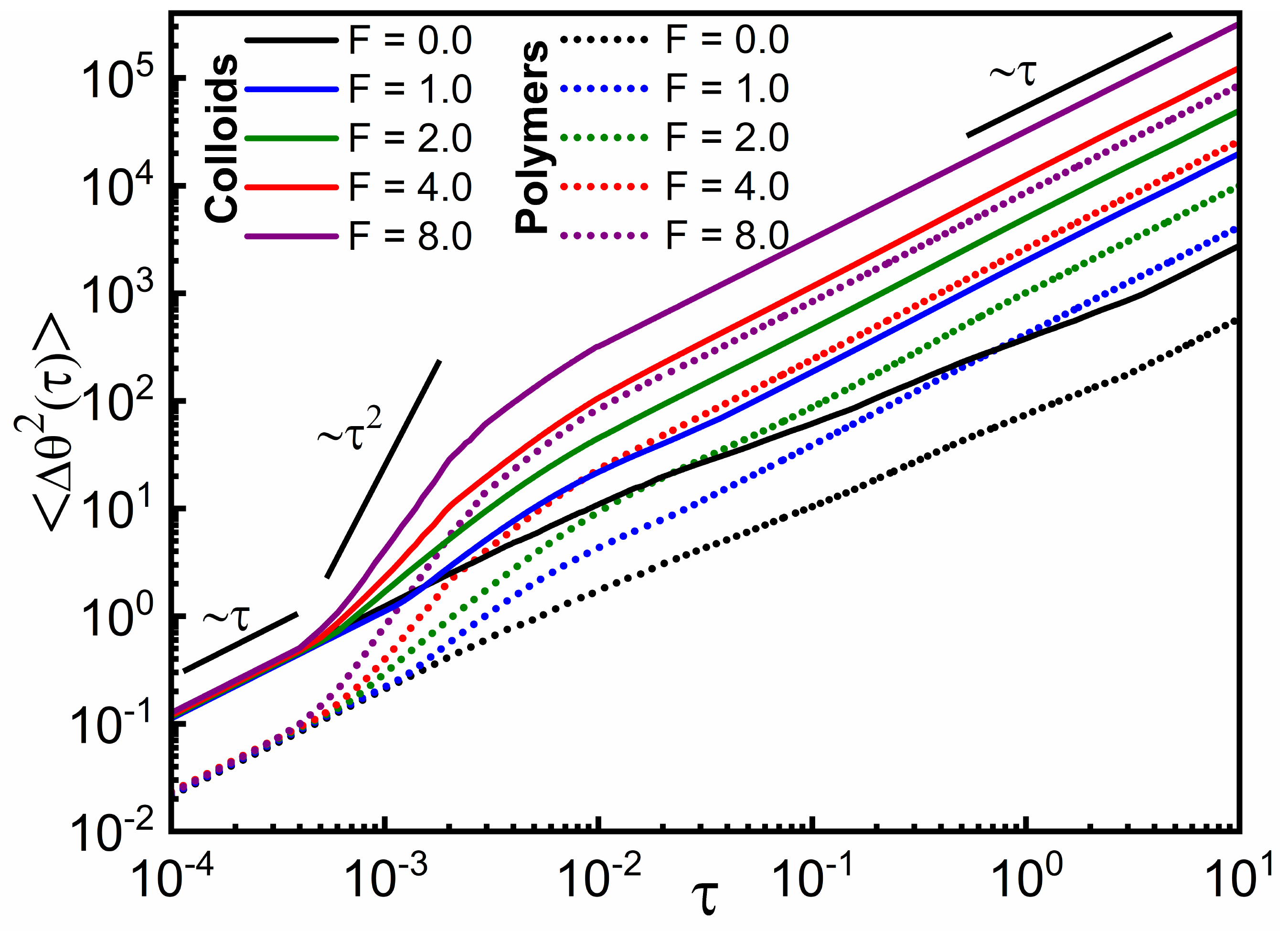} &
		\includegraphics[width=0.5\textwidth]{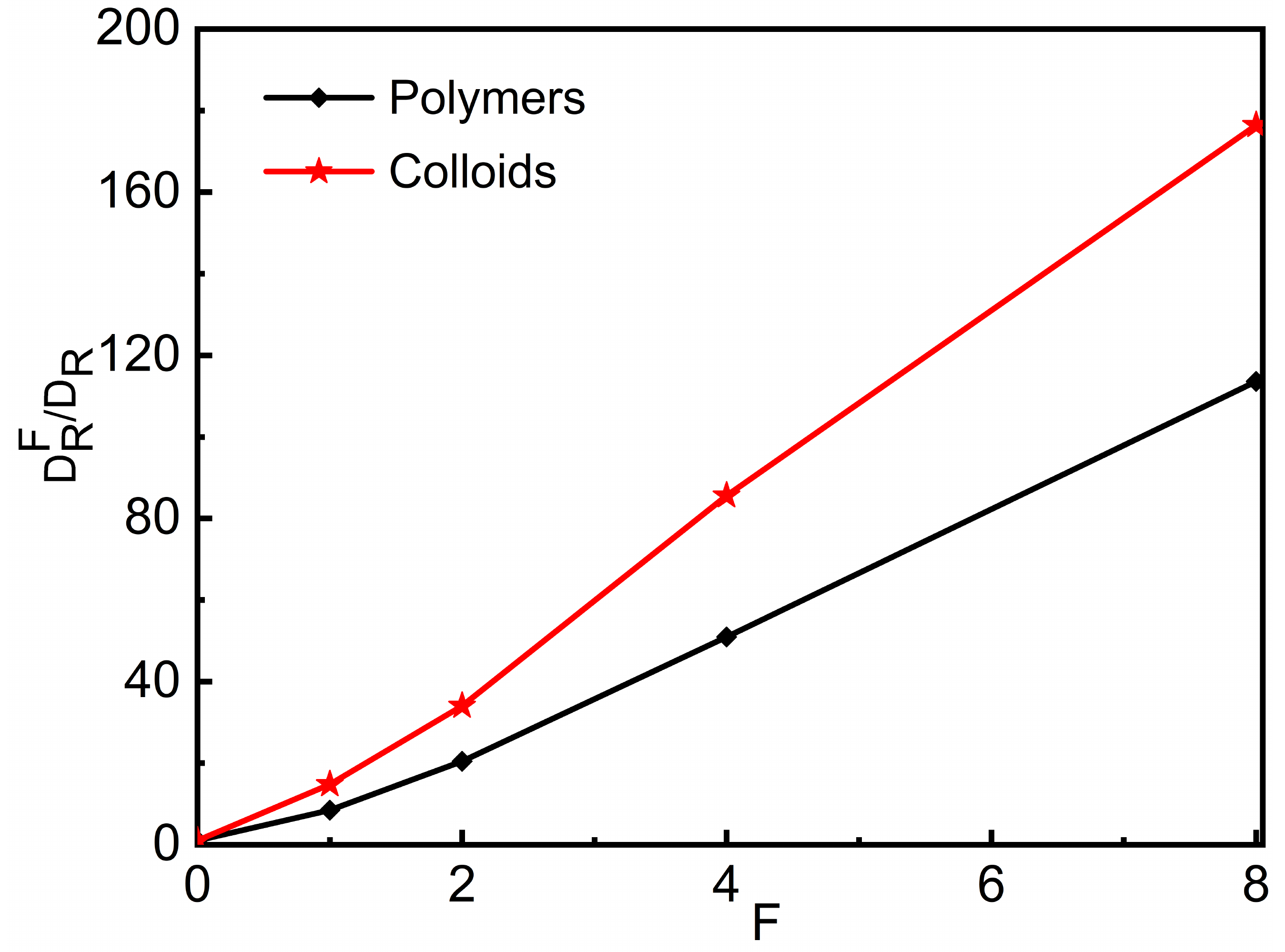} \\
	
		(a) & (b)  \\
		
	\end{tabular}
		\caption{\small Log-Log plot of $\langle \Delta \theta^{2} (\tau) \rangle$ vs $\tau$ for the self-propelled Janus probe in the purely repulsive polymeric environment (dotted lines) and in purely repulsive colloids (solid lines) for $\phi = 0.165$. The black solid and dotted lines represent the case for the passive Janus probe and (b) Normalized rotational diffusion coefficient $\frac{D_{R}^{F}}{D_{R}}$ for the same case.}
\label{fig:MSAD_WCA}
\end{figure}

\noindent In order to investigate whether viscoelasticity is essential for our observations, we remove the bonds connecting the monomers of the polymer and carry out the simulations for purely repulsive monomers, as well as for a mixture of attractive and repulsive monomers [see ESI, Movie-4].  Absence of connectivity eliminates the viscoelastic response of the medium and $\left<\Delta r_c^2(\tau)\right>$, $\left<\Delta \theta^2(\tau)\right>$ for the purely repulsive colloids show similar trends for a fixed self-propulsion force but a faster growth as compared to in repulsive polymers (Fig.~\ref{fig:TMSD_WCA},~\ref{fig:MSAD_WCA}). The normalized rotational diffusion coefficient $\left(\frac{D^F_R}{D_R}\right)$ of Janus probe for different $F$ shows that the diffusion is much faster in colloids compared to the Janus probe in polymeric medium {with crowder area fraction $\phi=0.165$} (Fig.~\ref{fig:MSAD_WCA}(b)). Where, $D^{F}_R$ is the rotational diffusion coefficient with self-propulsion $F$ in the presence of crowders and $D_R$ is the same with no self-propulsion. $D^{F}_R$ ($D_R$) is obtained by fitting the long time $\left<\Delta \theta^2(\tau)\right>$ with $4D^{F}_R\tau$ ($4D_R\tau$). Hence, we hypothesize that the origin of this enhancement is due to the additional torque arising from the activity-dependent interactions between the Janus probe and the sea of free passive particles so as in the case of polymers. Since the monomers are connected, the motion of individual polymer beads is constrained compared to that of the free colloidal beads. In a viscoelastic environment, the Janus probe is restricted inside cavities created by long chains leading to a diminution of diffusivity as compared to the colloid crowders. To establish this point further, we carry out simulations with frozen polymers and frozen colloids separately [see ESI, Movie-5, and Movie-6]. We see that in the case of frozen colloids, the enhancement of rotation is more than the frozen polymers (ESI, Fig.~5). However, in the presence of crowders, the translational and rotational dynamics of the Janus probe  slow down in general as compared to the free Janus particle. In a mixture of sticky and non-sticky beads, the self-propelled Janus with two different directions of self-propulsion show similar qualitative trends in $C_\textrm{v}(\tau)$ (Fig.~\ref{fig:VACF_1}(b)), $\left<\Delta r_c^2(\tau)\right>$ (ESI Fig.~6(a)) and $\left<\Delta \theta^2(\tau)\right>$ (ESI Fig.~7(a)) like in viscoelastic environment. In this case also, both $\left<\Delta r_c^2(\tau)\right>$ and $\left<\Delta \theta^2(\tau)\right>$ grow slowly (ESI Fig.~8) with increasing the size of the Janus probe like the case with polymers.

\subsection {Decoupling between translational and rotational diffusion}
\label{non_monotonous}
\begin{figure}
	\centering
	\begin{tabular}{cc}
		\includegraphics[width=0.5\textwidth]{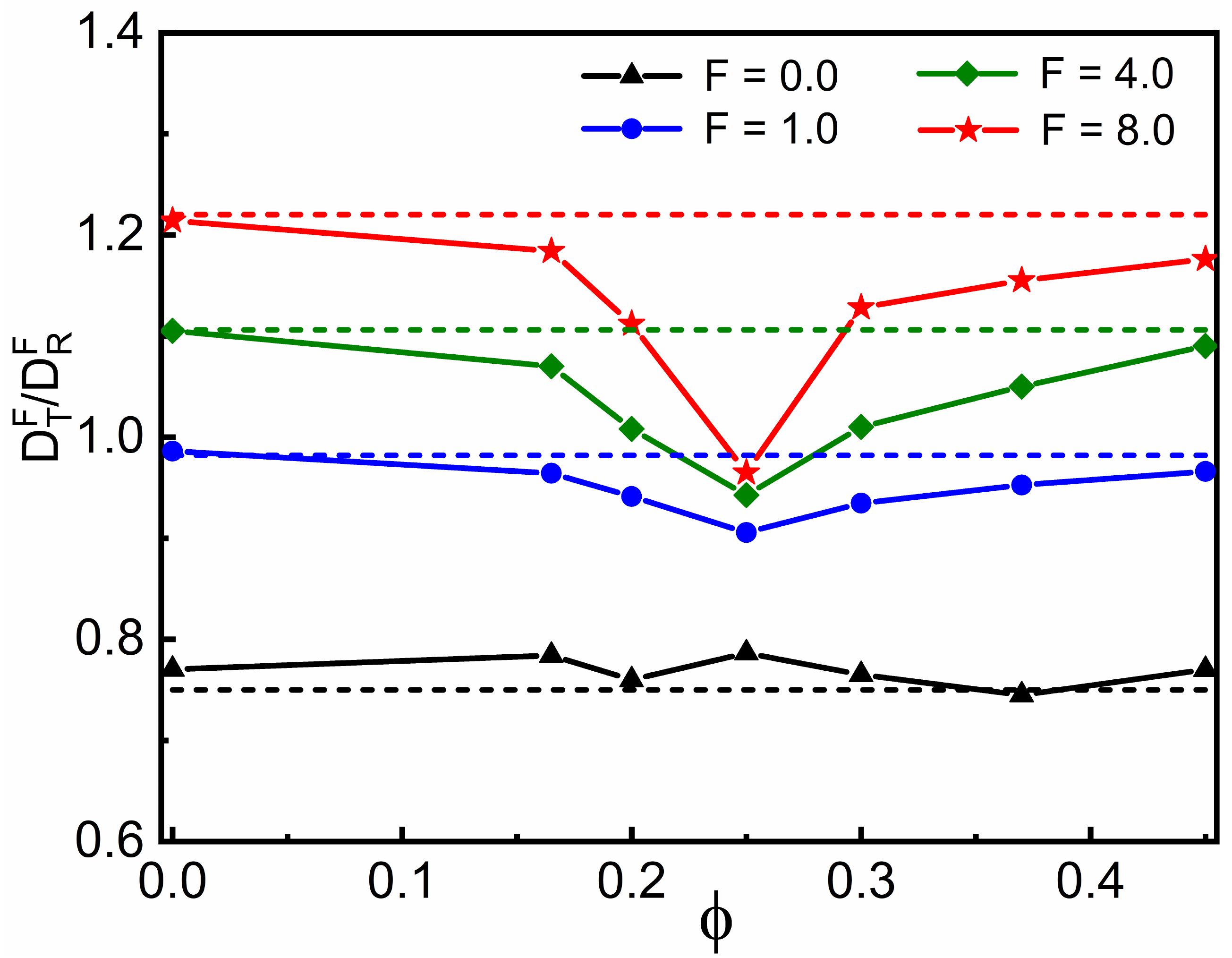} &
		\includegraphics[width=0.5\textwidth]{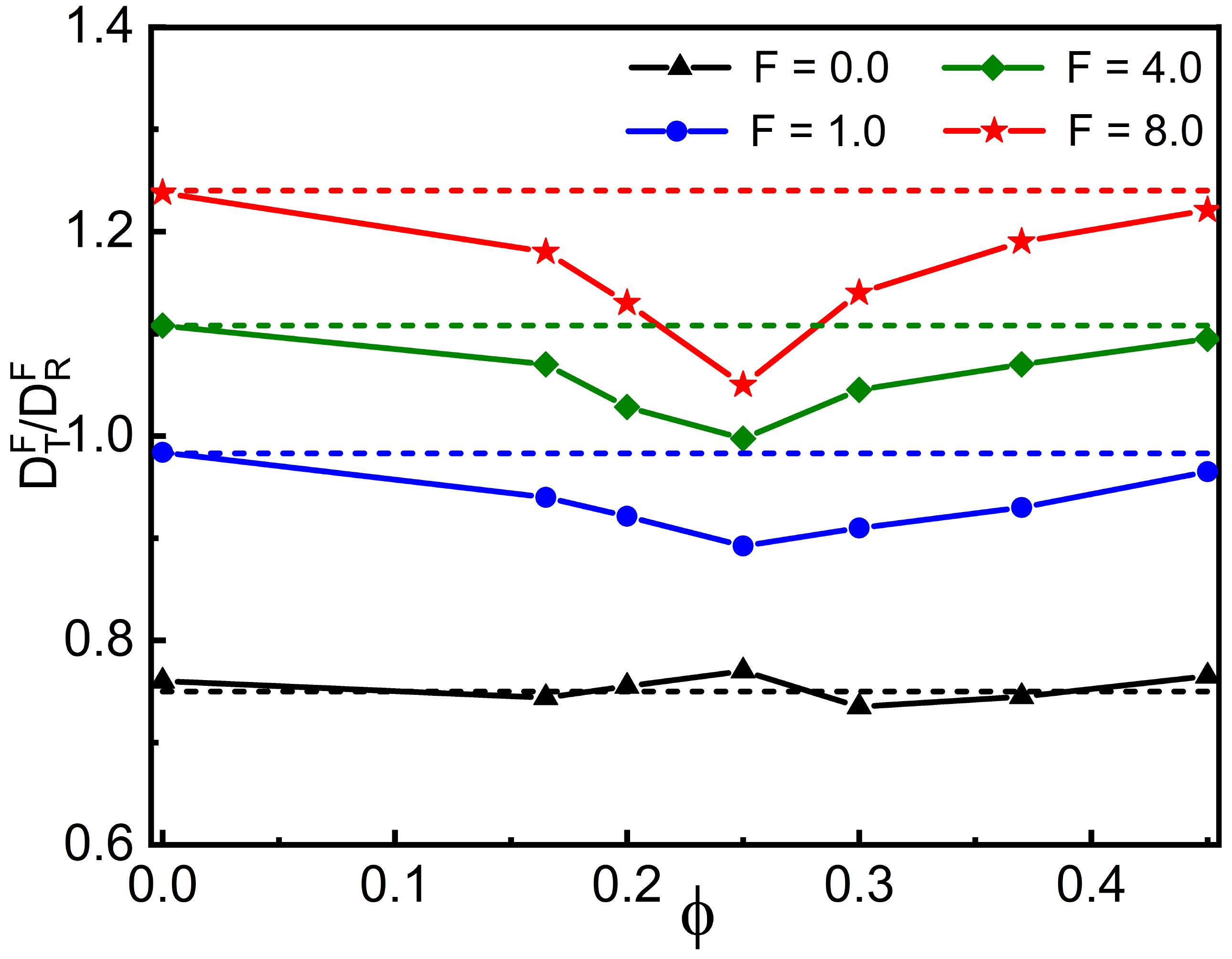} \\

		(a) & (b)  \\
		
	\end{tabular}
	\caption{\small The ratio $\frac{D^F_{T}}{D^F_{R}}$ for the self-propelled Janus probe in different area fraction of (a) polymers having sticky zones ($\epsilon = 2.0$) (b) binary mixture of attractive ($\epsilon = 2.0$) and repulsive colloids for different $F$. The dashed lines represent the value of $\frac{D^F_{T}}{D^F_{R}}$ for $\phi = 0$.}
\label{fig:Ratio_Diffusivity}
\end{figure}

\begin{figure}
	\centering
	\begin{tabular}{cc}
		\includegraphics[width=0.5\textwidth]{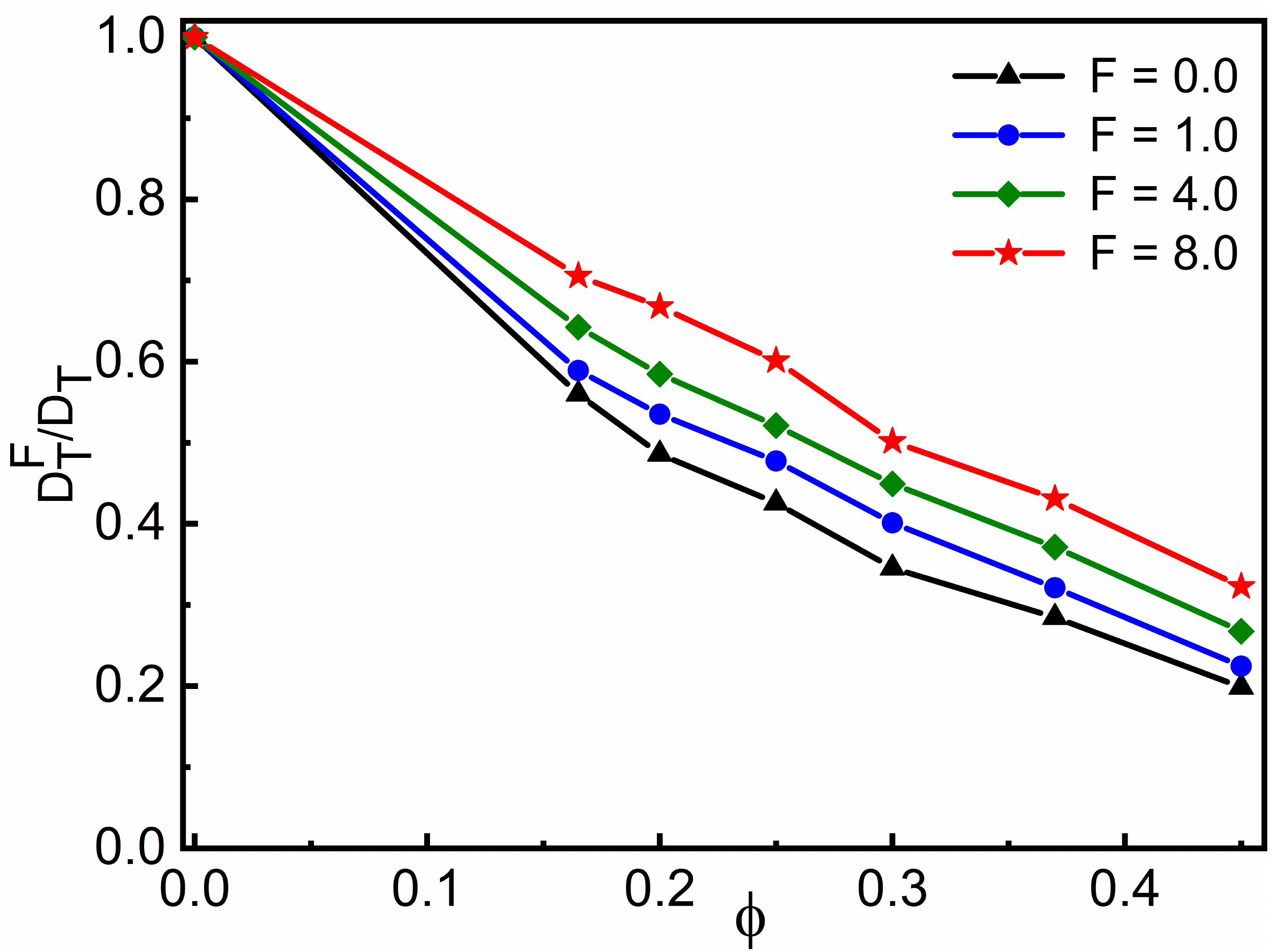} &
		\includegraphics[width=0.5\textwidth]{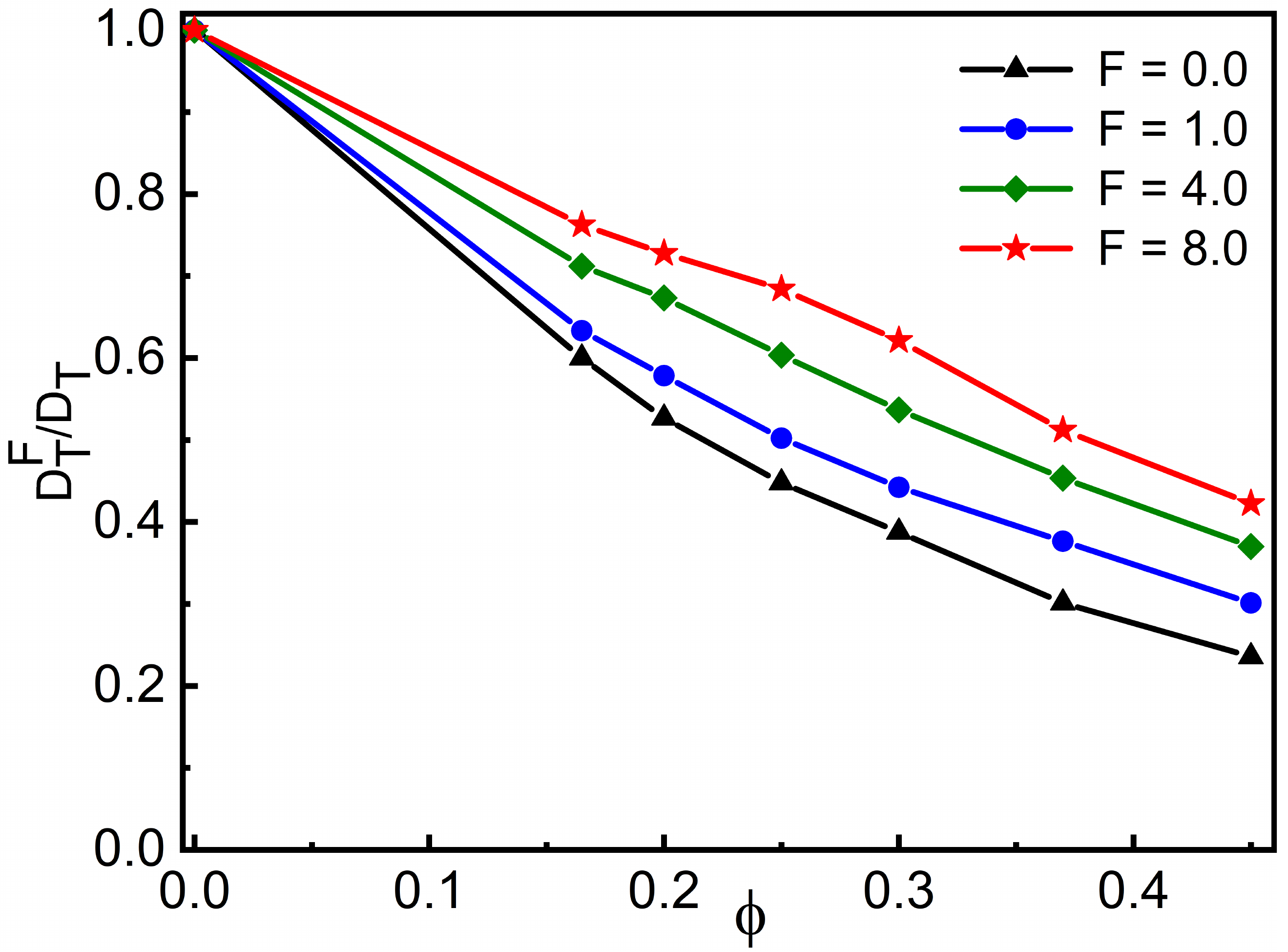} \\
		
		(a) & (b)  \\
		
	\end{tabular}
	\caption{\small The normalized translational diffusion coefficient $\left(\frac{D^F_T}{D_T}\right)$ for the self-propelled Janus probe in different area fraction of (a) polymers having sticky zones ($\epsilon = 2.0$) (b) binary mixture of attractive ($\epsilon = 2.0$) and repulsive colloids for different $F$.}
\label{fig:Translation_Diffusivity}
\end{figure}

\begin{figure}
	\centering
	\begin{tabular}{cc}
		\includegraphics[width=0.5\textwidth]{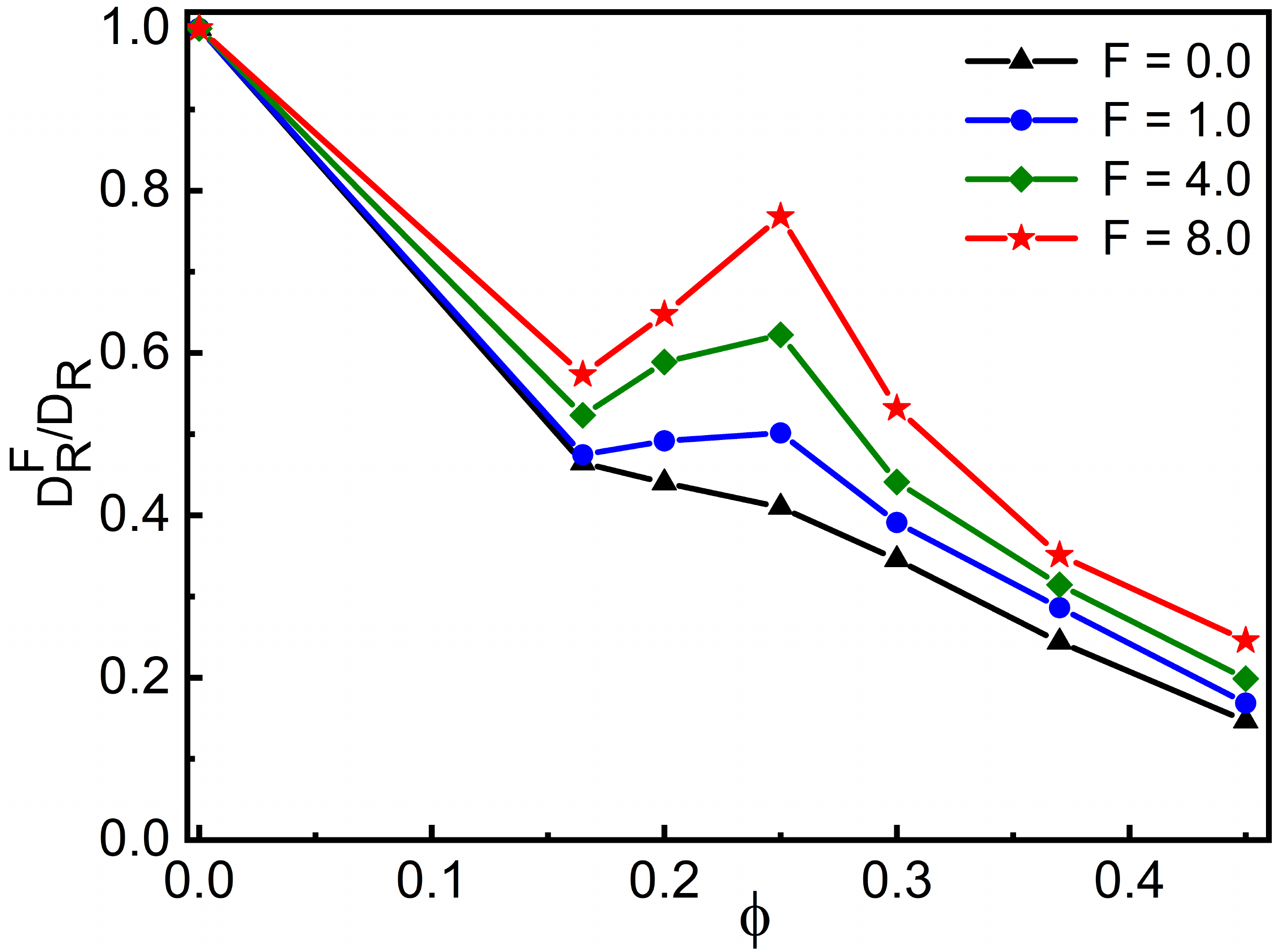} &
		\includegraphics[width=0.5\textwidth]{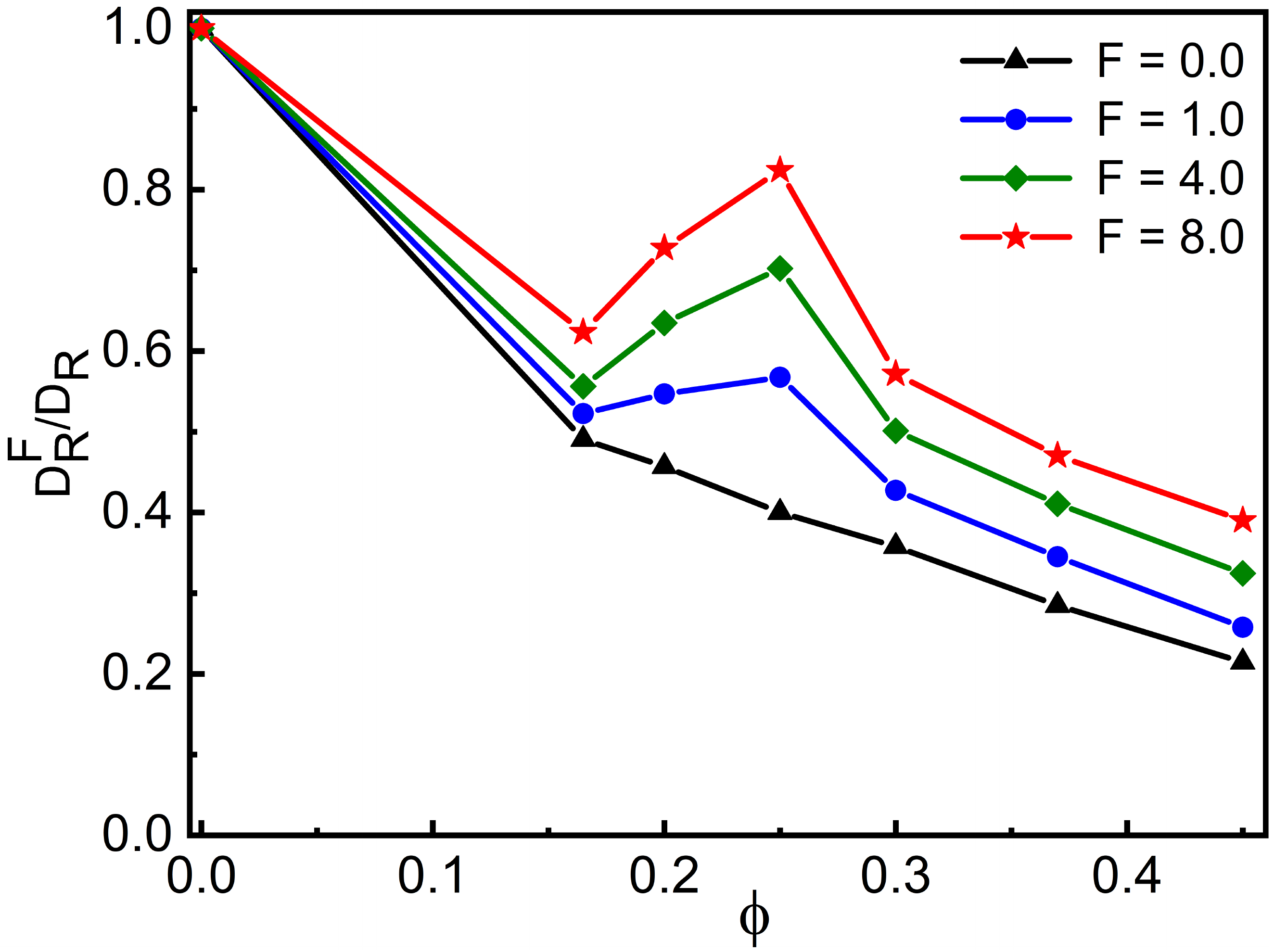} \\
		
		(a) & (b)  \\
		
	\end{tabular}
	\caption{\small The normalized rotational diffusion coefficient $\left(\frac{D^F_R}{D_R}\right)$ for the self-propelled Janus probe in different area fraction of (a) polymers having sticky zones ($\epsilon = 2.0$) (b) binary mixture of attractive ($\epsilon = 2.0$) and repulsive colloids for different $F$.}
\label{fig:Rotation_Diffusivity}
\end{figure}

\noindent The translational and rotational diffusion remain coupled as long as their ratio of the respective diffusion coefficients remains a constant. 
The rotation of a free Janus particle is purely governed by thermal diffusion, thus the ratio of translational and rotational diffusion coefficients ($\frac{D_T}{D_R}$) is a constant. However, any deviation of this ratio from the constant value should result from the decoupling of translational and rotational motions~\cite{edmond2012decoupling}. Spatial heterogeneity in the medium results in such a decoupling~\cite{zou2019coupling,makuch2020diffusion}. Thus, for a free self-propelled Janus particle, translational and rotational motions are always coupled.\\

\noindent We extract the values of effective translational and rotational diffusion coefficients of the Janus probe in the presence of viscoelastic crowders with sticky zones ($D_T^F$ and $D_R^F$, respectively), from the data shown in Fig.~\ref{fig:MSD_phi} [also see ESI, Fig.~9-12], for different values of $\phi$, {where, like in the case of rotational diffusion, $D_T^F$ ($D_T$) is obtained by fitting the long time $\left<\Delta r_c^2(\tau)\right>$ with $4D^F_T \tau$ ($4 D_T \tau$)}. In Fig.~\ref{fig:Ratio_Diffusivity}(a), we plot the ratio, $\frac{D^F_T}{D^F_R}$ as a function of $\phi$ for different $F$. In the passive case ($F=0$), $\left( \frac{D^F_T}{D^F_R} \right)$ is independent of $\phi$, indicating a coupling between translational and rotational diffusion {at} all densities. However, when $F > 0$, the diffusivity ratio shows a strong dependence of $\phi$, where it shows a minimum around $\phi \simeq 0.25$. This behavior becomes more pronounced for higher values of $F$. For high density $\phi \gtrsim 0.4$ the ratio approaches the free-Janus value (at $\phi \simeq 0$). Repeating the same analysis for non-viscoelastic crowders with the same number density provide qualitatively similar results (Fig.~\ref{fig:Ratio_Diffusivity}(b)). The origin of this non-monotonous behavior is evident from Fig.~\ref{fig:Translation_Diffusivity} and Fig.~\ref{fig:Rotation_Diffusivity}, where we separately plot the normalized diffusivity values ($\frac{D^F_T}{D_T}$ and $\frac{D^F_R}{D_R}$) as a function of $\phi$. While $D^F_T$ steadily decreases with increase in $\phi$, $D^F_R$ shows a non-monotonous behavior, showing a maximum around $\phi \simeq 0.25$. The possible reason for this non-monotonous behavior in $D_R^F$ is qualitatively described as follows. \\

 \noindent In a crowded environment, the Janus probe gets surrounded by its neighboring particles (polymers or colloids) and the mean free path of the Janus probe decreases with an increase in $\phi$ (and $F$). Since the rotation also changes the self-propulsion direction, the probe {with enhanced diffusivity} translates to a different location after rotation, where it encounters interactions from the surrounding particles. Each interaction event  induces a random rotation, which leads to an increase in $D^F_R$ with $F$ at a small $\phi$ (Fig.~\ref{fig:Rotation_Diffusivity}). However, such rotations are suppressed at even larger $\phi$, as the crowding does not provide sufficient free space for the probe to rotate. Thus, the non-monotonous behavior is possibly due to two competing effects; first due to the enhancement in rotation induced by probe-crowder interaction, and second due to the suppression of rotation at a large $\phi$. However, this non-monotonic behavior is less pronounced {or absent} for $D_T^F$ as a persistent translational motion gets suppressed with an increase in $\phi$. \\

\noindent We have fitted the data to the Eq.~\ref{eq:MSD_Analytical} for different area fractions of polymers or colloids, with $\tau_R$ and $v$ as the fitting parameters. We have found that, Eq.~\ref{eq:MSD_Analytical} fits with all the $\left<\Delta r_c^2(\tau)\right>$ curves for polymers as well as colloids (See ESI, Fig.~13(a) and Fig.~14(a)). From Table~1 in the ESI, it is evident that $v$  monotonically decreases with $\phi$, whereas $\tau_R$ exhibits a non-monotonic behavior. The non-monotonic behavior in angular diffusion (Fig.~\ref{fig:Rotation_Diffusivity}) is get reflected in the fitting parameter $\tau_R$, as $D^F_R=\frac{1}{\tau_R}$. Hence, the observed enhancement in angular diffusion is attributed to the effect of the collisions between the Janus probe and the surrounding particles.

\section{Conclusions}
\noindent In this paper, we have investigated the role of crowders in controlling the translational and rotational dynamics of a Janus probe. This Janus probe is either passive or self-propelled and the crowders are either polymers or colloids. In addition, crowders can have sticky or repulsive interactions with the probe. When the crowders are sticky,  the Janus nature of the probe becomes evident as then the direction of self-propulsion, whether it originates from the sticky or non-sticky face becomes one of the key factors. Our simulation results imply that there is an enhancement in the translational, as well as rotational diffusion on changing the passive probe to a self-propelled one, in the presence of crowding. Also, in the presence of crowders, the rotational diffusivities increase by orders of magnitude as a function of the self-propulsion force $F$. This dependence of rotational dynamics of Janus probes on self-propulsion is absent for free Janus particles. Moreover, the general trend of enhancement  observed in rotational diffusion is independent of the fact that whether the medium is viscoelastic or not. It is even more pronounced in a medium with no viscoelastic response. However, in the presence of crowders, the translational and rotational dynamics of the Janus probe, in general slow down as compared to the free Janus particle. Most importantly, we observe that the rotational motion of the Janus particle is decoupled from its translational motion at intermediate area fractions. This decoupling gets stronger with increasing activity. However, at high area fractions, the ratio of translational and rotational diffusivities approaches its $\phi = 0$ value. But for the passive Janus, translational and rotational diffusions are always coupled, irrespective of nature and the area fraction occupied by the crowders. As the passive Janus probe lacks the additional torque coming from the combined effect of self-propulsion and crowding. The self-propulsion results in frequent collisions with the crowders and subsequently generates an additional torque, responsible for faster rotation. For passive probes, collisions with crowders are less frequent and it has no mechanism to generate excess torque from the environment.   But, at high area fractions, the rotational motion of the self-propelled Janus slows down due to the substantial steric hindrance created by the crowders. \\

\noindent In this work, we have considered a moderately dense system, up to a crowder area fraction of $0.45$. Recent experiments \cite{klongvessa2019active,klongvessa2019nonmonotonic} have shown that in the case of dense sediment of Brownian particles, the tagged particle in the glassy state has a non-monotonic dynamics with self-propulsion velocity. In other words, the translational relaxation time when plotted against an effective temperature, it first increases and then decreases. Such non-monotonicity \cite{szamel2015glassy,chakiescape2020} is absent in our case. Our simulations always predict a monotonic increase in translational and rotational mean square displacements and hence the diffusivities with the self-propulsion force, for a given area fraction of crowders. This makes us explore a future problem, probe dynamics in active glassy systems and to investigate the existence of any non-monotonicity of translational or rotational motion of the probe with the activity for a given density.

\section{Acknowledgements}
\noindent We thank Murugappan Muthukumar for insightful discussions. LT thanks UGC for a fellowship. SC thanks DST Inspire for a fellowship. $RC^{\ast}$ acknowledges SERB (Project No. SB/SI/PC-55/2013) and IRCC-IIT Bombay (Project No. RD/0518-IRCCAW0-001) for funding and $RC^{\dagger}$ acknowledges the financial support from SERB, India via the grants SB/S2/RJN-051/2015 and SERB/F/10192/2017-2018. We acknowledge the SpaceTime-2 supercomputing facility at IIT Bombay for the computing time.


\end{document}